\newcommand{\gusten}{G\"{u}sten}
\newcommand{\kms}{km s${}^{-1}$}
\newcommand{\al}{$\alpha$}
\newcommand{\ab}{$\sim$}
\newcommand{\yz}{Yusef-Zadeh}
\newcommand{\p}{$\pm$}
\newcommand{\ad}{$\alpha$,$\delta$$_{(J2000)}$}
\newcommand{\beam}{beam$^{-1}$}
\newcommand{\x}{$\times$}
\newcommand{\T}{T$_e^*$}
\newcommand{\pc}{pc$^{-1}$}

\documentclass[preprint2]{aastex}

\begin{document}

\title{A VLA H92$\alpha$ Recombination Line Study of the Arched Filament \\H II Complex Near the Galactic Center}   

\author{Cornelia C. Lang\altaffilmark{1,2}, W. M. Goss\altaffilmark{2}, Mark Morris\altaffilmark{1}}
\altaffiltext{1}{Department of Physics \& Astronomy, 8371 Math Sciences Building, University of California, Los Angeles, CA 90095-1562; CCL's current address: Astronomy Program, LGRT B-517O, University of Massachusetts, Amherst, MA 01003, email: clang@ocotillo.astro.umass.edu}
\altaffiltext{2}{National Radio Astronomy Observatory, Box 0, Socorro, NM 87801}
 
\begin{abstract}

The Very Large Array has been used at 8.3 GHz in the DnC and CnB array
configurations to carry out an H92\al~recombination line study (at 8.3 GHz) of the
ionized gas in the Arched Filaments H II complex, which defines the
western edge of the Galactic center Radio Arc. The Arched
Filaments consist of a series of curved filamentary ridges which
extend over 9\arcmin\x6\arcmin~(22 \x~16 pc) and are intersected in
two places by linear, nonthermal, magnetic filaments. The
H92\al~observations cover the entire Arched Filaments region using a
four-field mosaic, with an angular resolution of 12\farcs8
\x~8\farcs10; an additional higher resolution (3\farcs6 \x~2\farcs7)
field was imaged in the SW portion of the H II complex. High
resolution continuum images are also presented. The H92\al~line
properties of the ionized gas (line-to-continuum ratio, FWHM line
width, \T) are consistent with photoionization from hot stars, and
consistent with the physical properties of other Galactic center H II regions. The LTE electron temperatures vary only slightly across the entire extent of
the source, and have an average value of 6200 K. The velocity field
is very complex, with velocities ranging from +15 \kms~to $-$70 \kms~and the majority of velocities having negative values. Large velocity
gradients (2$-$7 \kms~\pc, with gradients in some regions $>$ 10
\kms~\pc) occur along each of the filaments, with the velocities
becoming increasingly negative with decreasing distance from the
Galactic center.  The negative velocities and the sense of the
velocity gradients can be understood in terms of the orbital path of
the underlying molecular cloud about the Galactic center. The magnitudes of the velocity gradient are consistent with the cloud residing on an inner, elongated orbit which is due to the Galaxy's stellar bar, or with a radially infalling cloud.
The ionization of the Arched Filaments can be accounted for completely by the massive Arches stellar cluster, which consists of $>$ 150 O-stars and produces a few \x~10$^{51}$ photons s$^{-1}$. This cluster is likely to be located 10$-$20 pc from the Arched Filaments, which can explain the uniformity of ionization conditions in the ionized gas.     
   
\end{abstract}


\section{Introduction}
Several of the
most unusual H II regions in the Galaxy are found within 30 pc of the Galactic center. High resolution radio observations
made with the Very Large Array (VLA) of the National Radio Astronomy
Observatory\footnotemark\footnotetext{The National Radio Astronomy Observatory is a facility
of the National Science Foundation, operated under a cooperative
agreement with the Associated Universities, Inc.} over the past 15 years have
revealed the remarkable filamentary morphology of H II regions such
as the Sickle and Pistol (\yz~\& Morris
1987a; Lang, Goss \& Wood 1997), the Arched Filaments (\yz~1986), and SgrA
West (Ekers et al. 1983; Schwarz, Bregman, \& van Gorkom 1989, Roberts \& Goss 1991, 1993). 
The largest and most prominent of these regions is the ``Arched Filament'' H II
complex, comprised of a series of curved, narrow ridges of radio emission which define the western edge of the well-known Galactic center Radio Arc (Yusef-Zadeh et
al. 1984). The Arched Filaments extend for 9\arcmin~$\times$
6\arcmin~(or 22 $\times$ 15 pc at the assumed Galactic center distance of 8.0
kpc (Reid 1993)) and are located 10\arcmin~(25 pc) in projection from
the center of the Galaxy, SgrA$^*$. 
The thermal nature of these filamentary ridges was first
revealed by recombination line observations (Pauls et al. 1976; Pauls
\& Mezger 1980; Yusef-Zadeh 1986; Yusef-Zadeh, Morris \& van Gorkom 1987). 
In addition to the peculiar morphology of these thermal filaments, the recombination line studies have shown that the kinematics of the Arched
Filaments are also very striking, with large velocity gradients along their lengths and predominantly negative velocities in a positive  
velocity quadrant of the Galaxy, i.e., counter to Galactic rotation
for circular orbits. A large molecular cloud complex was discovered in
CS (J=2$-$1) at the position of the Arched Filaments and extends
southward over 16\arcmin~(or 40 pc) near SgrA. This cloud exhibits
emission over a range of velocities similar to those of the ionized
gas (i.e., 5 to $-$55 \kms) and is therefore known as the ``$-$30
\kms~cloud'', its name representing the average velocity of the gas in this
cloud (Serabyn \& \gusten~1987).

In addition to these unusual properties of the Arched Filaments, the
surrounding interstellar environment is remarkable
for several reasons. 
First, at their location near the Galactic center, the strong differential gravitational forces of this region are likely to influence both the morphology and kinematics of the interstellar gas
within the inner kiloparsec (\gusten~\& Downes 1980; Serabyn \& \gusten~1987) and may well
play an important role in understanding the Arched Filaments. 
Second, the Arched Filaments are intersected by two prominent systems of
non-thermal filaments (NTFs). These NTFs are unique
to the central 250 pc of the Galaxy; the long (up to 50 pc) and narrow
($<$0.1 pc) synchrotron filaments show strong linearly polarized
radio emission and have magnetic field orientations aligned with their long
axes (Yusef-Zadeh \& Morris 1987b; Yusef-Zadeh, Wardle \& Parastaran
1997; Lang et al. 1999ab). The NTFs are understood as evidence for a large-scale
poloidal magnetic field which pervades the central 250 pc of the Galaxy
(Morris 1994). The NTFs in the Radio Arc intersect the northern edge of the Arched Filaments, and to the southwest,
the Northern Thread NTF crosses the southern end of two of the Arched Filaments
(Morris \& Yusef-Zadeh 1989, hereafter MYZ; Lang et al. 1999b). The nature of the
intersection between the thermal gas and NTFs remains a major outstanding issue in understanding the interstellar
medium at the Galactic center. The origin and acceleration of
relativistic particles in the NTFs may be due to magnetic reconnection
at positions where the NTFs are interacting with associated ionized
and molecular gas (Serabyn \& Morris 1994).

Finally, one of the most exceptional, massive stellar clusters in the
Galaxy is located at the eastern edge of the Arched Filaments. The
brightest sources in this cluster were first
revealed by medium resolution, near-infrared
observations (Nagata et al. 1995; Cotera et al. 1996). The near-infrared
spectroscopy of Cotera et al. (1996) showed that 13 of the stars can be classified as
highly-evolved massive stars such as Wolf-Rayet (WR) and Of stellar
types. The ionizing flux generated by these 13 stars may provide part of the
ionization of the Arched Filaments. 
More recent, high resolution observations of the ``Arches
Cluster'' made at
near-infrared wavelengths at the W.M. Keck Observatory and with the NICMOS camera on the 
{\it Hubble Space Telescope (HST)} show that it is
very densely packed, harboring more than 150 O-stars (Serabyn et
al. 1998; Figer et al. 1999). In addition, Figer et al. (1999)
estimate that the cluster has a total mass of \ab10$^{4}$ M$_{\sun}$,
with an age of only \ab2 Myr. A number of far-infrared (FIR)
studies have revealed strong FIR line and continuum emission arising
from the Arched Filaments (Genzel et al. 1990; Erickson et al. 1991;
Morris et al. 1995; Davidson et al. 1994; Colgan et al. 1996). 

Before this luminous stellar cluster was discovered, initial explanations for heating of the Arched Filaments relied
on shocks between the $-$30 \kms~molecular cloud and the interstellar
medium in the Galactic center (Bally et al. 1988; Hayvaerts, Norman
\& Pudritz 1988). In addition, MYZ proposed that the source of
heating of the ionized filaments was MHD-induced ionization resulting
from the large relative velocity between the molecular cloud and the
strong, ambient magnetic field. So far, none of these models has been able to
account for all aspects of the radio and FIR observations.
In particular, Colgan et al. (1996) reported that the FIR
luminosity, [OIII] line fluxes, and low electron densities cannot be
explained by either shocks or MHD models for ionization. Instead,
these authors have concluded that the Arched Filaments must be
uniformly photoionized by a distribution of massive stars.  

In this paper, new VLA observations of the H92\al~recombination lines arising
from the Arched Filaments are presented. These observations were carried out in
order to make a detailed study of the ionized gas in this unusual source. These data cover the entire
region of the Arched Filaments using a four-field mosaic, with higher spatial resolutions (2$-$12\arcsec) than the previous VLA
H110$\alpha$ study of Yusef-Zadeh (1986), which was centered only on
the western filaments with a resolution of 22\arcsec. The goals of
these observations are to understand: (1) the peculiar morphology of
the Arched Filaments, (2) the complicated kinematics and velocity
field, (3) the interaction of the ionized gas with the 
massive and luminous stellar cluster and
(4) the nature of intersections between NTFs and ionized gas. $\S$2 provides a summary of the observations
and data reductions; results from the 8.3 GHz continuum images are
presented in $\S$3; results from the H92$\alpha$ line observations are
presented in $\S$4, and $\S$5 includes a discussion of the morphology,
kinematics and ionization of the Arched Filaments. 

\section{Observations and Data Reduction}
VLA continuum and recombination line observations of the Arched Filaments were made at 8.3 GHz
in the DnC and CnB array configurations. Details of
the observations are summarized in Tables 1 and 2. Calibration and editing
were carried out using the {\it AIPS} software of NRAO. Line-free channels were used to determine the continuum, and continuum
subtraction was done in the (u,v) plane using the {\it AIPS} task UVLSF. The four fields observed in DnC configuration were mosaicked using a maximum entropy deconvolution algorithm (mosmem) in the {\it MIRIAD} software package.  The H92$\alpha$ line and corresponding continuum mosaics were imaged with natural weighting to
achieve the best possible signal-to-noise ratio in the recombination line.
An additional continuum mosaic was created with uniform weighting in order to obtain higher spatial resolution. 
The southeastern field (Arches1) was also observed in the CnB array,
and these data were combined with
the corresponding DnC observations using the {\it AIPS} task DBCON. The combined uv-data were imaged
using the {\it AIPS} task IMAGR for both the 8.3 GHz continuum and the H92\al~line with natural weighting. A higher resolution 8.3 GHz
image of a portion of this field was made with uniform weighting to highlight the fine-scale structure. The parameters of all images
discussed in this paper are summarized in Table 3. The recombination
line analysis was done using {\it GIPSY}, the Groningen Image Processing
SYstem (van der Hulst et al. 1992).  Spatially-integrated, continuum-weighted
line profiles were made for selected regions of the Arched Filaments
using PROFIL, and Gaussian models were fitted to these averaged profiles using
PROFIT. Single-component Gaussian functions were also fitted to the line
data for each pixel having a signal-to-noise ratio $>$ 4.

\begin{deluxetable}{lccccc}
\tablecaption{Summary of Observations}
\tablewidth{0pt}
\tablehead{
\colhead{Field}& 
\colhead{Date}&
\colhead{Array}&
\multicolumn{2}{c}{Phase Center}&
\colhead{Integration Time}\\
\cline{4-5}
\colhead{}&
\colhead{}&
\colhead{}&
\colhead{$\alpha$ (J2000)}&
\colhead{$\delta$ (J2000)}&
\colhead{(hours)}}
\tablecolumns{5}
\startdata
Arches 1&Feb 1999&DnC&17 45 35.0&$-$28 50 00&8\\
``&Nov 1998&CnB&''&''&28\\
Arches 2&Feb 1999&DnC&17 45 47.4&$-$28 50 00&8\\
Arches 3&Feb 1999&DnC&17 45 47.4&$-$28 47 18&8\\
Arches 4&Feb 1999&DnC&17 45 35.0&$-$28 47 18&8
\enddata
\end{deluxetable}

\begin{deluxetable}{lc}
\tablewidth{0pt}
\tablecaption{Parameters of the H92$\alpha$ Line Observations}
\tablehead{\colhead{Parameter}&\colhead{H92$\alpha$~Data}}
\tablecolumns{2}
\startdata
H92$\alpha$~Rest Frequency&8309.382 MHz\\
LSR Central Velocity     &$-$40 \kms \\
Total Bandwidth  &12.5 MHz (45 \kms)\\
Number of Channels      &63\\
Channel Separation        &195.3 kHz (7.0 \kms)\\
Velocity Coverage&448 \kms\\
Flux Density Calibrator &3C286\\
Bandpass Calibrator& NRAO 530 (1730-130)\\
Phase Calibrator        &1751-253
\enddata
\end{deluxetable}

\begin{deluxetable}{llcccc}
\tablewidth{0pt}
\tablecaption{Summary of Arched Filaments Images}
\tablehead{
\colhead{Image}&
\colhead{}&
\colhead{Resolution}&
\colhead{PA}&
\colhead{RMS noise}&
\colhead{Weighting}}
\tablecolumns{6}
\startdata
Arches Mosaic&{\it continuum-high}&7\farcs78 $\times$
6\farcs61&$-$1\fdg0&0.9 mJy \beam~&uniform\\
&{\it continuum-low}&12\farcs84 \x~8\farcs10&$-$3\fdg4&1.5 mJy \beam~&natural\\
&{\it line}&12\farcs84 $\times$ 8\farcs10&$-$3\fdg4 &0.8 mJy \beam~&natural\\
Arches1&{\it continuum-high}&2\farcs26 $\times$ 1\farcs58&64\fdg2&0.2 mJy \beam~&uniform\\
&{\it continuum-low}&3\farcs61 \x~2\farcs66&44\fdg6&0.5 mJy \beam~&natural\\
&{\it line}&3\farcs61 $\times$ 2\farcs66&44\fdg6&0.3 mJy \beam~&natural
\enddata
\end{deluxetable}

\section{Continuum Images at 8.3 GHz}
Figure 1 is a schematic of the sources near the Arched
Filament H II complex which are discussed in this paper and depicted in
Figures 2 and 3. The 8.3 GHz continuum mosaic of the Arched Filaments is shown in greyscale and contours in Figures 2 and 3, with a resolution
of 7\farcs8 $\times$ 6\farcs6, PA=$-$1.0. Both images have been
corrected for primary beam attenuation. The 8.3 GHz continuum image is
very similar to the 1.4 and 4.8 GHz continuum images of MYZ. The Arched Filaments are comprised of two pairs of filamentary structures, the ``eastern'' and ``western''
filaments: E1, E2, W1, and W2 (after MYZ). The filaments are long and very narrow structures, extending up to
\ab9\arcmin~(22 pc) in the north-south direction, with widths of
only 20\arcsec~(0.8 pc) on average. The four Arched Filaments cover an area of
\ab6\arcmin~(15 pc) from east to west, but most of the emission is
concentrated in the narrow ridges, which have an areal filling factor
of \ab10\%. The similarity in the curvature of all four Arched Filaments is especially striking, and gives the appearance of
the western half of a set of concentric circles, the apparent center of
which would be located 1$-$2\arcmin~to the East.     

\begin{figure}[t!]
\plotone{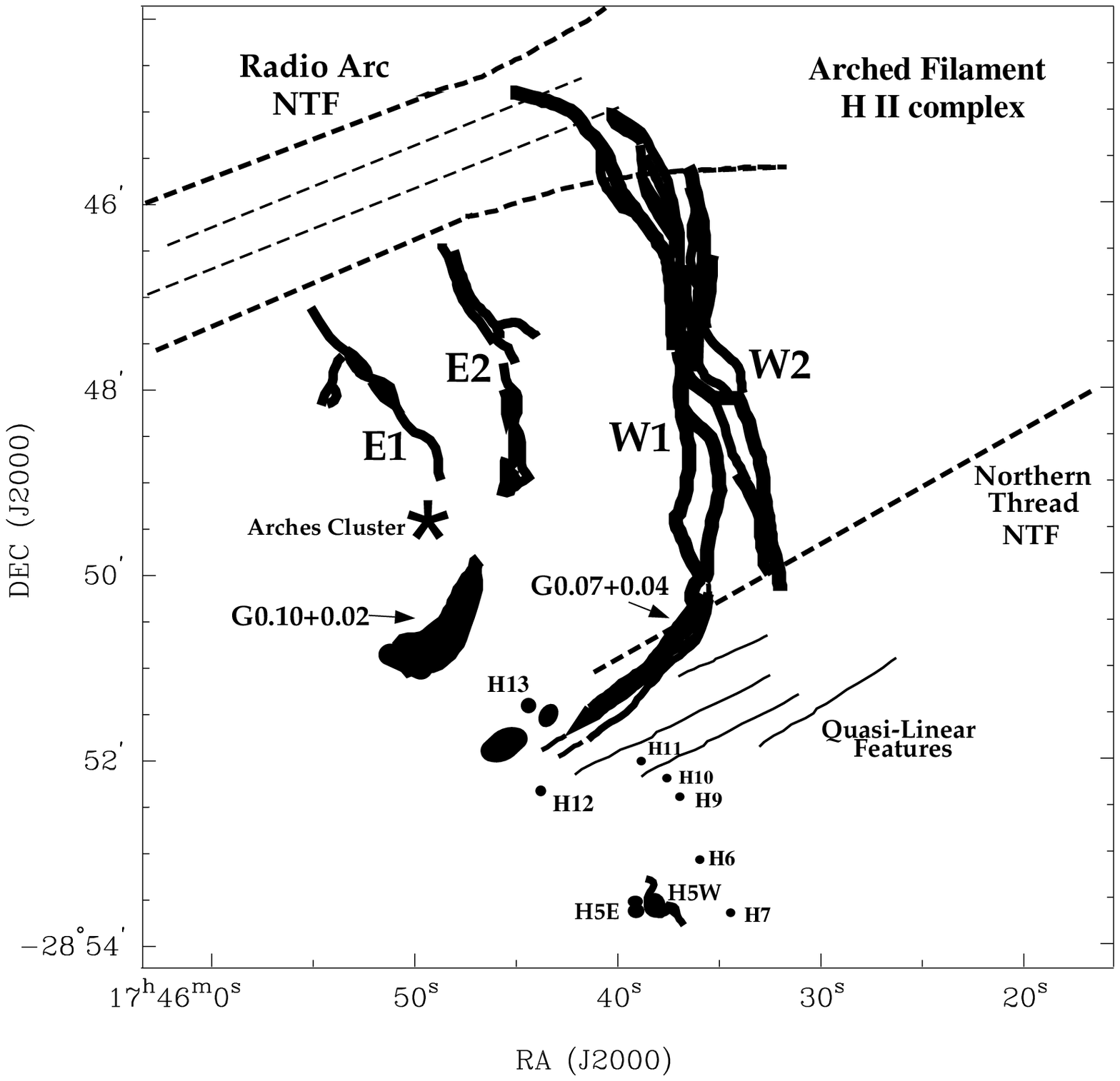}
\caption{Schematic representation of the Arched Filaments region as
shown in Figure 2.}
\end{figure}

\begin{figure}[t!]
\plotone{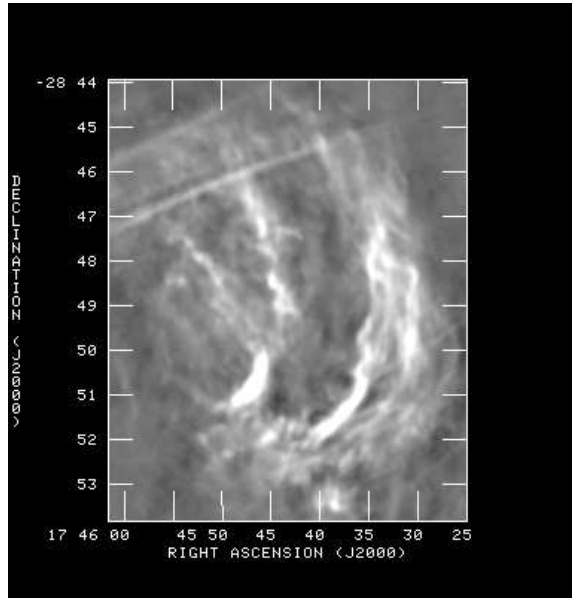}
\caption{VLA 8.3 GHz continuum image of the Arched Filaments and Radio Arc
(to the North) and Northern Thread (faintly
visible to the West) NTFs. This image was mosaicked from four fields using uniform weighting, and has been corrected for primary beam
attenuation. The resolution is 7\farcs78 $\times$ 6\farcs61,
PA=$-$1\fdg0, with an rms level of 0.9 mJy \beam.  Corresponding contours
are shown in Figure 3. At the Galactic center, 1\arcmin~corresponds to
2.5 pc.}
\end{figure}

\begin{figure}[t!]
\plotone{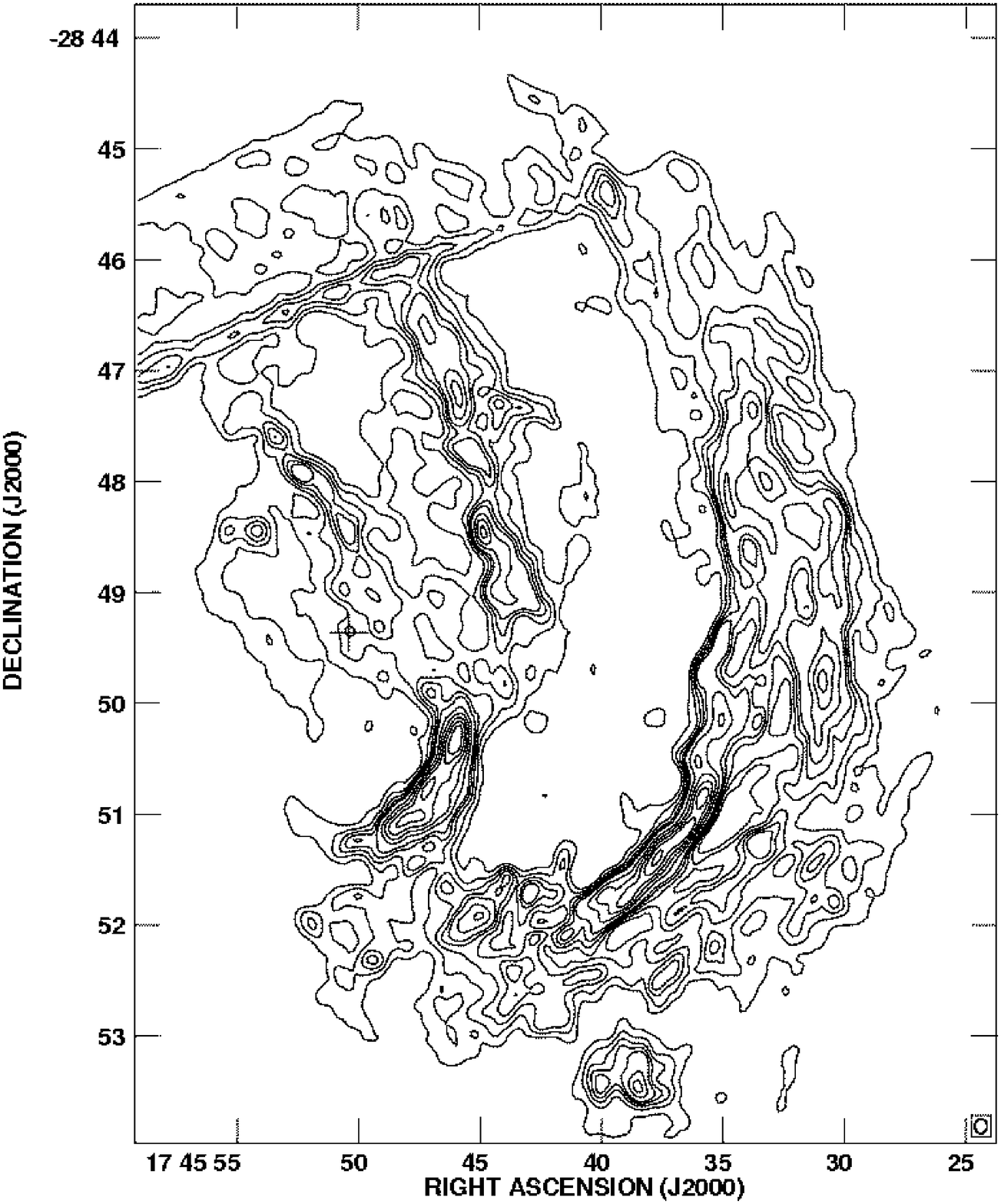}
\caption{8.3 GHz continuum contours which correspond to the image
shown in greyscale in Figure 2, with a resolution of 7\farcs78 $\times$
6\farcs61, PA=$-$1.0. The contours represent 5, 10, 15, 20, 25, 37.5, 50, 62.5,
75, 87.5, 100 mJy \beam. The cross represents the position of the
Arches stellar cluster.}
\end{figure}

Across the Arched Filaments, the brightness is unevenly
distributed, and has a diffuse and tenuous nature.  
In particular, along W1 and W2, the filaments appear to be comprised of
multiple filamentary strands, with the edges of these regions markedly brighter than the central portions of the filaments.
The brightest peaks, known as G0.07+0.04 and G0.10+0.02, have relatively uniform brightness over their  60$-$90\arcsec~extents, whereas the brightness over the rest of the filaments is much less uniform.  
Although it appears that G0.10+0.02 may be connected to E1 or E2, there is a
dramatic decrease in brightness between G0.10+0.02 and the southern
extent of the E1 and E2 filaments (at \ad=17 45 47.0, $-$28 49 12).
In this region, the emission drops to 15 mJy \beam, whereas to
the North (in E1), the intensity is $\geq$ 25 mJy \beam, and the intensity in G0.10+0.02 exceeds 85 mJy \beam. However, this is not the case for the G0.07+0.04 region and the W1 filament. Along W1, the emission does not vary as substantially as in the E1, E2, and G0.10+0.02 region. Moving northward along W1, the intensity increases fairly constantly from 50 to 80 mJy \beam~for \ab3\farcm5, followed by a decrease in intensity to \ab50 mJy \beam~starting at \ad=17 46 35.0, $-$28 50 00. There are no dramatic changes or discontinuities in the intensity along W1, whereas G0.10+0.02 has a sharper boundary that separates it from the E1 and E2 filaments.  Therefore, in this paper, G0.10+0.02 will be considered a separate source from E1 and E2, whereas the W1 filament will include G0.07+0.04.  

The Radio Arc NTFs (to the North) and Northern Thread NTF (to the SW)
are apparent in Figure 2 where they intersect (in projection) the
Arched Filaments. Yusef-Zadeh \& Morris (1988) have pointed out that the morphology and discontinuity of the NTFs where they
meet the Arched Filaments suggests that there may be a physical connection. On the western side of this image, the Northern Thread NTF is just perceptible where it
crosses, in projection, both W1 and W2. 
A collection of diffuse, extended sources is located to the south of the W2 filament.
Several of these features are nearly linear and are oriented parallel to
the NTFs, including the adjacent Northern Thread NTF. These ``quasi-linear'' features were first
pointed out by MYZ. 

Based on the continuum flux density at 8.3 GHz, physical parameters of
the ionized gas in the Arched Filaments have been derived using the formulation of Mezger \& Henderson (1967) and the
corrections to this formulation (Viallefond 1991). In using these
formulae, we assume a uniform density, spherical, ionization-bounded H II region with \T=6200 K, and Y$^+$=0.06 (see $\S$4). Although the Arched Filaments are obviously not spherically-symmetric, we divide the region into segments which can be better approximated by a spherical H II region in order to derive the standard parameters for comparison with other H II regions. The ionization-bounded assumption is also valid for the regions we are considering, as the radio emission arises in the the Arched Filament complex from the ionization-bounded side of the nebula. Part of the nebula (to the east of the Arches cluster) is likely to be density-bounded, as the distribution of underlying molecular material falls off completely (Serabyn \& \gusten~1987) and there is no detectable free-free emission arising from this region.

\begin{figure}[t!]
\plotone{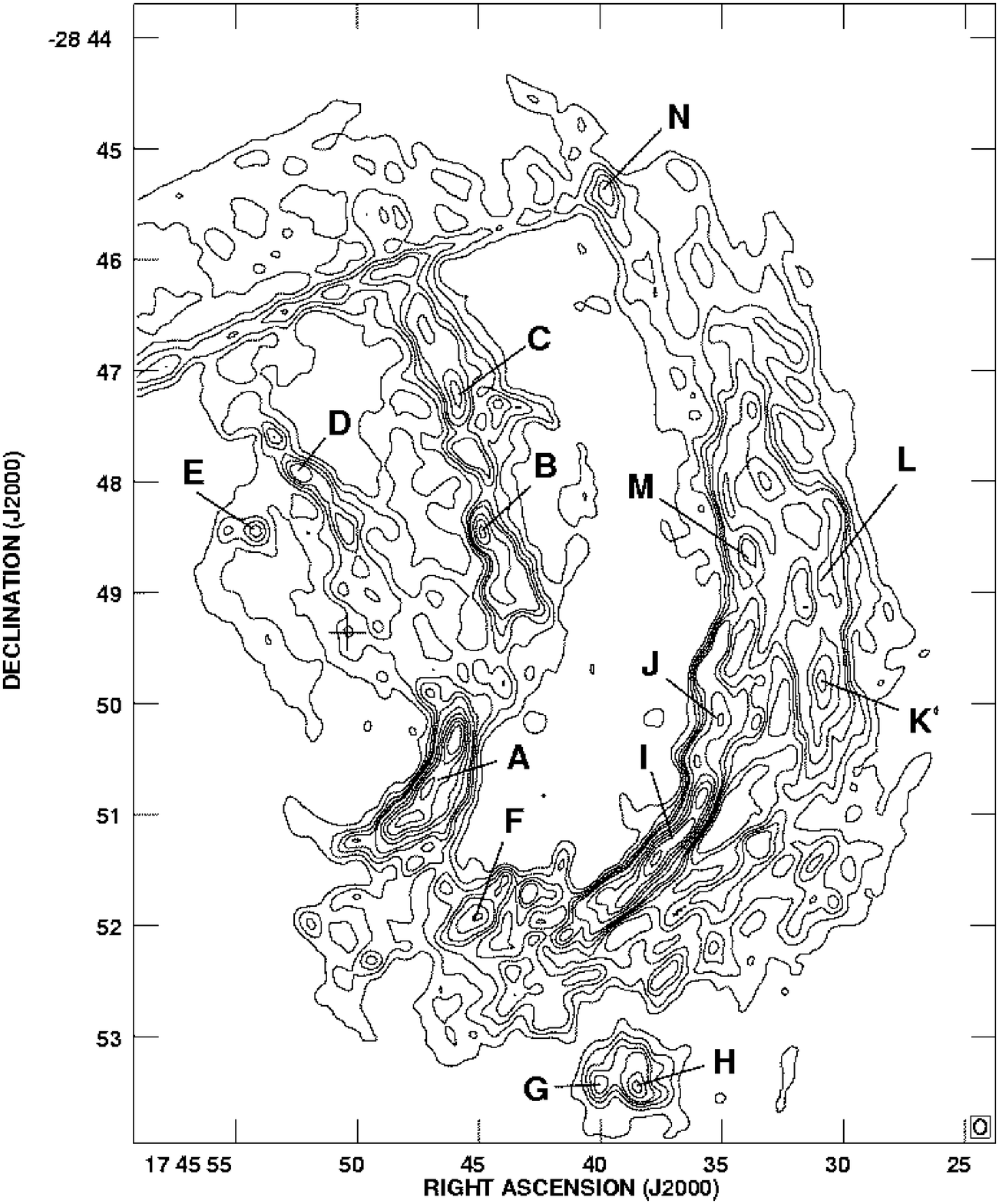}
\caption{8.3 GHz continuum image as shown in Figure 3. A-N represent
the regions for which physical parameters of the ionized gas were derived from the radio continuum (see Table 4).}
\end{figure}

Table 4 lists the derived parameters for these regions: total flux density (S$_\nu$), measured angular size, linear size of the equivalent sphere, electron density (n$_e$),
emission measure (EM), ionization parameter (U), mass of ionized hydrogen (M$_{HII}$), number of
Lyman continuum ionizing photons, (N$_{Lyc}$), and the 8.3 GHz continuum optical depth ($\tau_c$). 

\begin{deluxetable}{lccccccccc}
\tablewidth{0pt}
\tablecaption{Physical Quantities Derived from the 8.3 GHz Radio Continuum}
\tablehead{
\colhead{Source}&
\colhead{S$_{8.3}$}&
\colhead{Size}& 
\colhead{Radius\tablenotemark{*}}& 
\colhead{n$_e$} & 
\colhead{EM}&
\colhead{U}&
\colhead{M$_{HII}$} & 
\colhead{N$_{Lyc}$} & 
\colhead{$\tau_{c}$}\\
\colhead{}&
\colhead{(Jy)}&
\colhead{(\arcsec)}&
\colhead{(pc)}&
\colhead{(cm$^{-3}$)}&
\colhead{(pc cm$^{-6}$)} &
\colhead{(pc cm$^{-2}$)}&
\colhead{(M$_{\sun})$}&
\colhead{(s$^{-1})$}& 
\colhead{(8.3 GHz)}}
\tablecolumns{10}
\startdata
A\tablenotemark{a}& 2.0	&100\x30&1.5 &	290&2.6$\times$10$^{5}$	&68&106&1.6$\times$10$^{49}$&0.002\\
B& 0.9&70\x30&1.3 &250&1.7$\times$10$^{5}$&52&55&7.2$\times$10$^{48}$ &0.001\\
C &0.9&80\x20&1.1 &310& 2.2$\times$10$^{5}$&52&45&4.5$\times$10$^{48}$ &0.002\\
D& .05&12\x12&0.3 &460&1.4$\times$10$^{5}$&20&2&3.6$\times$10$^{47}$&0.001\\
E& 0.1&20\x20&0.6 &360&1.4$\times$10$^{5}$&29&6&1.2$\times$10$^{48}$&0.001\\
F& 0.4&30\x30&0.8 &340&1.9$\times$10$^{5}$&41&21&2.6$\times$10$^{48}$&0.001\\
G\tablenotemark{b}& 0.2&15\x20&0.5 &520&2.7$\times$10$^{5}$&32&6&1.5$\times$10$^{48}$&0.002\\
H\tablenotemark{c}& 0.3&20\x20&0.6 &510&3.0$\times$10$^{5}$&37&9&2.3$\times$10$^{48}$&0.002\\
I\tablenotemark{d}& 2.2	&120\x30&1.7&270&2.5$\times$10$^{5}$&72&132&1.6$\times$10$^{49}$&0.002\\
J& 0.9&80\x20&1.1&300&2.2$\times$10$^{5}$&52&45&7.0$\times$10$^{48}$&0.002\\
K& 1.2&70\x30&1.3&290&2.2$\times$10$^{5}$&57&63&1.0$\times$10$^{49}$&0.002\\
L& 0.4&40\x30&1.0&250&1.3$\times$10$^{5}$&40&24&3.3$\times$10$^{48}$&0.001\\
M& 1.6&100\x50&2.0&180&1.2$\times$10$^{5}$&63&140&1.3$\times$10$^{49}$&0.001\\
N& 0.7	&25\x25	&0.7&310&1.4$\times$10$^{5}$&33&11&1.4$\times$10$^{48}$&0.001
\enddata
\tablenotetext{*}{Assuming a distance of 8.0 kpc to the Galactic center.}
\tablenotetext{a}{Also known as radio continuum peak G0.10+0.12 from MYZ; far
infrared peak in both line and continuum (Erickson et al. 1991).}
\tablenotetext{b}{Known H II region H5(W); 6 cm flux density=770 mJy from Zhao
et al. 1993.}
\tablenotetext{c}{Known H II region H5(E); 6 cm flux density=330 mJy from Zhao
et al. 1993.}
\tablenotetext{d}{Also known as radio continuum peak G0.07+0.04 from MYZ.}
\end{deluxetable}

\begin{deluxetable}{lccccc}
\tablewidth{0pt}
\tablecaption{Continuum Sources near the Arched Filaments}
\tablehead{\colhead{Source}&
\multicolumn{2}{c}{Position}&
\multicolumn{2}{c}{Flux Density (mJy)}&
\colhead{Spectral}\\
\cline{2-3}
\cline{4-5}
\colhead{}&
\colhead{$\alpha$ (J2000)}&
\colhead{$\delta$ (J2000)}&
\colhead{3.6 cm\tablenotemark{*}}&
\colhead{6 cm\tablenotemark{**}}&
\colhead{Index}}
\tablecolumns{6}
\startdata
H6	&17 45 36.6	&$-$28 53 09	&23.3\p3.0	&19.3\p2.0&+0.3\\
H7	&17 45 35.0	&$-$28 53 35	&23.8\p3.0	&16.0\p2.0&+0.7\\
H9	&17 45 37.6	&$-$28 52 42	&6.7\p0.5	&9.0\p0.5&$-$0.5\\
H10	&17 45 37.8	&$-$28 52 34	&4.2\p0.5	&5.8\p1.0&$-$0.6\\
H11	&17 45 38.8	&$-$28 52 31	&4.3\p0.8	&4.4\p0.8&+0.0\\
H12	&17 45 43.7	&$-$28 52 26	&3.2\p0.4	&5.5\p2.2&$-$0.9\\
H13	&17 45 43.9	&$-$28 51 33	&2.7\p0.2	&5.9\p1.0&$-$1.4
\enddata
\tablenotetext{*}{From Figure 5}
\tablenotetext{**}{From Figure 3 in MYZ}
\end{deluxetable}

\begin{figure}[t!]
\plotone{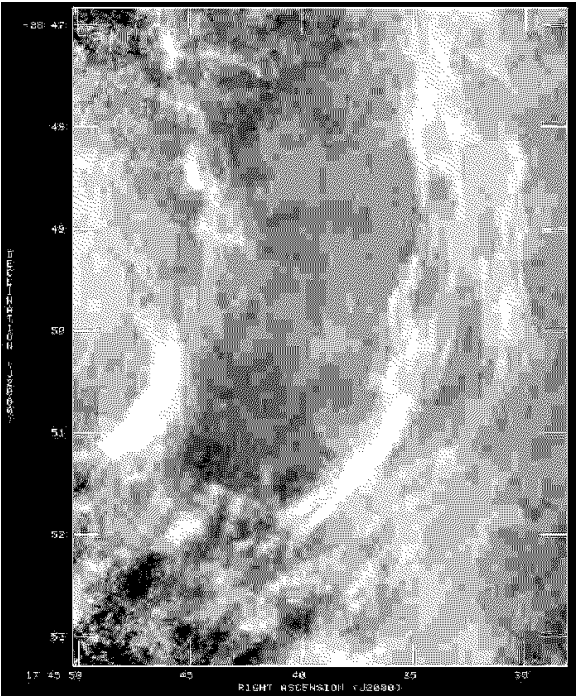}
\caption{High-resolution VLA 8.3 GHz continuum image of the southwestern
field of the Arched Filaments, shown in greyscale, which includes the emission peaks G0.07+0.04
and G0.10+0.02. This image was made from the combined CnB and DnC
array data, and has a resolution of 2\farcs26 $\times$
1\farcs58, PA=64\fdg2. The rms noise level level is 0.2 mJy \beam, and
this image has been corrected for primary beam attenuation.}
\end{figure}

Figure 5 shows a higher resolution 
(2\farcs3~\x~1\farcs6, PA=64\arcdeg) 8.3 GHz continuum image of a
single field in the southeastern portion of the Arched
Filaments. As in Figure 2, W1 and W2 appear comprised of multiple
filamentary strands and narrow ridges of projected width $<$ 3\arcsec~(0.125 pc). G0.07+0.04 is prominently bifurcated at an orientation nearly
perpendicular to the Galactic plane. Several of the point-like sources to the south of G0.07+0.04 were catalogued by Yusef-Zadeh (1986) as H5-H7. Four additional sources appear in this
image, and following the above nomenclature, they are labelled as H9-12 (see Figure 1) and the parameters are listed in Table 5. 

\section{H92\al~Recombination Line Observations}

Spatially-integrated, continuum-weighted H92\al~profiles were made over
each of the four Arched Filaments in order to characterize
their global H92\al~properties. For these profiles, G0.10+0.02 was assumed
to be a member of the E1 filament. The profiles are shown in
Figure 6, and properties of the Gaussian fits to these profiles (line-to-continuum ratio, T$_l$/T$_c$; central
LSR velocity, V$_{LSR}$; and FWHM line width, $\Delta$V) are given in Table 6. Single component profiles
were fitted to the H92$\alpha$ line in the E2, W1, and W2 filaments; region
E1 was fitted with a
double peaked profile. A possible detection of the He92\al~line was made in E2,
and the intensity ratio of helium to hydrogen, Y$^+$, was calculated to
be 0.04\p0.015. The central velocities in the W1 and W2 filaments (V=$-$27 and $-$44 \kms)
are more negative than those in E1 and E2 (V=$-$12 \kms), and the double
profile in the E1
filament suggests that the velocity field in this region may be
complex and comprised of multiple components. The line
widths of the integrated profiles in the Arched Filaments range from 27$-$38 \kms. Line widths for integrated regions within other Galactic center H II regions, such as G0.18-0.04, G0.15-0.05, and SgrA West were found to be much broader (typically \ab50 \kms), although smaller integrated regions in these regions have an average line width of \ab30 \kms~(Lang et al. 1997; Roberts \& Goss 1993). The increased line widths for the larger integrated regions in all cases can be attributed to a combination of the large velocity gradients across each filament and the large areas over which the profiles were integrated.  

\begin{figure}[t!]
\plotone{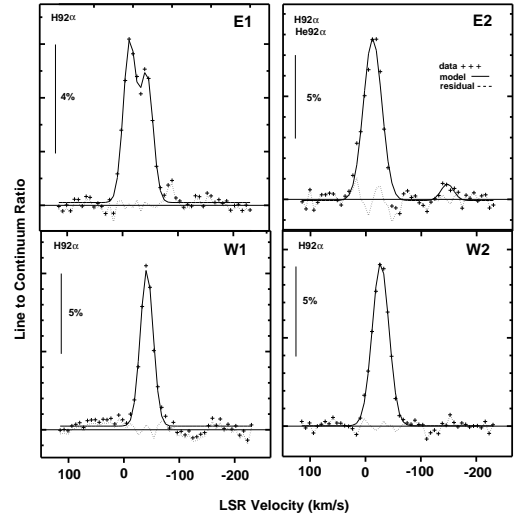}
\caption{Spatially integrated, continuum-weighted profiles for the
four filaments in the Arched Filaments - E1, E2, W1 and W2. The line
to continuum ratio in units of percent (values represented by vertical
bars) is plotted against central LSR velocity. Crosses represent the data points, solid lines show the model Gaussian fit, and dotted lines represent the
residuals.}
\end{figure}

\begin{deluxetable}{lcccc}
\tablewidth{0pt}
\tablecaption{Gaussian Properties of Integrated H92$\alpha$ Lines in
E1, E2, W1 \&~W2 Filaments}
\tablehead{
\colhead{Region}&
\colhead{T$_{l}$/T$_{c}$}& 
\colhead{V$_{LSR}$}&
\colhead{$\Delta$V}&
\colhead{T$_{e}^*$}\\
\colhead{No.}&
\colhead{}&
\colhead{(\kms)} &
\colhead{(\kms)} &
\colhead{(K)}} 
\tablecolumns{5}
\startdata
E1&0.06\p0.006&$-$12.9\p0.9&27.1\p1.9&6800\p700\\
...&0.05\p0.006&$-$43.8\p1.2&26.4\p2.3&...\\
E2&0.08\p0.005&$-$12.3\p0.7&38.2\p1.6&6000\p400\\
W1&0.10\p0.005&$-$41.2\p0.4&27.1\p0.9&6700\p400\\
W2&0.09\p0.003&$-$27.3\p0.3&35.2\p0.8&5800\p300
\enddata
\end{deluxetable}

\clearpage

\begin{figure}[t!]
\plotone{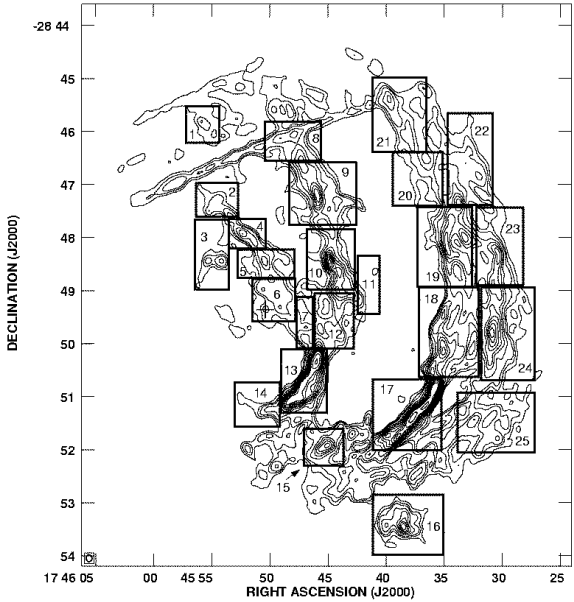}
\caption{Boxes superimposed on the 8.3 GHz continuum image (as shown
in Figure 3) of the Arched Filaments represent regions 1-25, over which the H92$\alpha$ line emission was integrated to produce the profiles shown in Figures 8-10.}
\end{figure}

The H92\al~line was also sampled on smaller scales at 25 positions
within the Arched Filaments, representing the well-defined emission
complexes illustrated in Figure 7. These 25 H92\al~profiles, also
spatially-integrated and continuum-weighted, are presented in Figures 8-10, and the corresponding Gaussian properties of these profiles are listed in Table 7. In order to examine the spatial variation in the H92$\alpha$ line
properties in further detail, single-component Gaussians
were fit to each pixel having a flux density above a 4$\sigma$ level. The resulting spatial distribution of line amplitude, FWHM line width, and LTE electron
temperature (derived from the measured values of T$_l$/T$_c$ and
$\Delta$V), are presented in Figures 11-13.  
 
The H92$\alpha$ recombination line flux densities have peak values in the range of 3 to 25 mJy \beam. The strongest H92$\alpha$ emission
(25 and 23 mJy \beam) occurs at the continuum peaks G0.07+0.04 and G0.10+0.02. 
As shown in Figure 11, the H92$\alpha$ line emission
follows the continuum emission closely, including the substructure and
multiplicity of filaments apparent along the length of W1.    
The northern boundary of the H92$\alpha$ emission in the Arched
Filaments coincides with the
NTFs in the Radio Arc. The emission abruptly declines toward the
southernmost NTF in the Radio Arc, with the exception of an H92$\alpha$ line arising from a small region in
the Radio Arc north of E2 (Profile 1). 
To the South, the H II region known as H5 (E \&
W) (Profile 16) represents the southernmost source of H92$\alpha$ line
emission in our data; this source has been studied in detail by Zhao
et al. (1993).

Detections of the He92$\alpha$ line have been made at the
2$-$3$\sigma$ level in several regions of the Arched Filaments (Profiles 7, 9, 16 \& 19). 
Values of Y$^+$ range from 4$-$8\% in these regions, consistent with other
radio recombination line studies of Galactic center H II regions,
which typically show Y$^+$\ab5\% (Mehringer et al. 1993; Roberts \&
Goss 1993; Lang et al. 1997).  The only known enhancement of
He92$\alpha$ in this region is the detection of Y$^+$=14\p6\% in
portions of the Pistol nebula (Lang et al. 1997), thought to be
metal-enriched ejecta from an earlier evolutionary stage of the Pistol star
(Figer et al. 1998).  

\begin{deluxetable}{lccccc}
\tablewidth{0pt}
\tablecaption{Gaussian Properties of the H92$\alpha$ Lines in Figures 8-10}
\tablehead{
\colhead{Region}&
\colhead{T$_{l}$/T$_{c}$}& 
\colhead{V$_{LSR}$}&
\colhead{$\Delta$V}&
\colhead{T$_{e}^*$}&
\colhead{T$_D$}\\
\colhead{No.}&
\colhead{}&
\colhead{(\kms)} &
\colhead{(\kms)} &
\colhead{(K)}&
\colhead{(K)}} 
\tablecolumns{6}
\startdata
1       &0.09\p0.01      &$-$20.9\p1.1     &22.1\p2.5&6800\p700\tablenotemark{*}&10500\p1200\\
2       &0.15\p0.01    	&$-$29.9\p0.7      &22.4\p1.6&5500\p700&10800\p800\\
3       &0.12\p.009    	&$-$39.1\p0.6      &28.6\p1.6 &5400\p500&17700\p1000\\ 
4       &0.16\p0.01    	&$-$20.0\p0.4       &15.8\p1.0 &7100\p800&5400\p400\\
5       &0.08\p.007    	&$-$8.4\p0.7      &29.4\p1.8 &7500\p800&18700\p1200\\
6       &0.11\p.008    	&$-$18.0\p0.4      &17.5\p1.0 &8900\p900&6600\p400\\
7       &0.08\p0.01      &$-$17.0\p1.4	   &32.0\p0.9&7000\p1000&22100\p700\\ 
8       &0.07\p0.005     &$-$26.6\p1.1      &35.0\p2.0&5000\p500\tablenotemark{*}&26500\p1500\\
9       &0.12\p.006    	&$-$19.8\p0.4	   &26.8\p1.0&5700\p400&15300\p600\\
10      &0.08\p.003   	&$-$3.4\p0.5 	&41.0\p1.2&5600\p300&36300\p1800\\
11	&0.17\p0.02  	&$-$33.0\p0.7	&16.7\p1.7&6400\p900&6000\p600\\
12	&0.07\p.004  	&+4.3\p0.6	&36.0\p1.7&5800\p800&28000\p1400\\
...	&0.04\p.005	&$-$39.2\p0.8	&18.9\p1.9&...&...\\
13	&0.09\p.005	&$-$28.1\p0.6	&42.4\p1.6&5000\p300&39000\p2000\\
14	&0.06\p.005  	&$-$1.0\p0.8	&27.6\p1.9&5600\p800&16500\p1100\\
...	&0.04\p.005	&$-$43.9\p1.0	&25.5\p2.5&...&...\\
15	&0.12\p0.01	&$-$25.7\p0.6	&21.4\p1.2&6900\p600&10000\p600\\
16	&0.08\p.007  	&$-$34.4\p1.0	&35.8\p2.4&6300\p600&28000\p1900\\
17	&0.12\p.007	&$-$43.6\p0.4	&22.6\p0.9&6700\p500&11000\p500\\
18	&0.12\p.006	&$-$36.8\p0.4	&25.2\p0.8&6000\p400&14000\p500\\
19	&0.09\p.004	&$-$26.4\p0.4	&37.4\p1.1&5500\p500&30000\p1200\\
20	&0.08\p.004	&$-$28.2\p0.7	&42.6\p1.6&5500\p300&39200\p1500\\
21	&0.10\p0.01	&$-$25.7\p0.6	&21.7\p1.5&6400\p700\tablenotemark{*}&10200\p700\\
22	&0.10\p0.01	&$-$8.6\p0.7	&25.8\p1.8&6900\p800&14400\p1000\\
23	&0.10\p.005	&$-$17.3\p0.5	&34.4\p1.2&5400\p300&26000\p1000\\
24	&0.13\p.006	&$-$27.0\p0.4	&25.9\p0.8&5500\p300&14500\p500\\
25	&0.10\p.008	&$-$39.6\p0.7	&28.4\p1.7&6400\p600&17000\p1000
\enddata
\tablenotetext{*}{Corrected for the 40\% non-thermal continuum due to the NTFs in the Radio Arc.}
\end{deluxetable}

\begin{figure}[t!]
\plotone{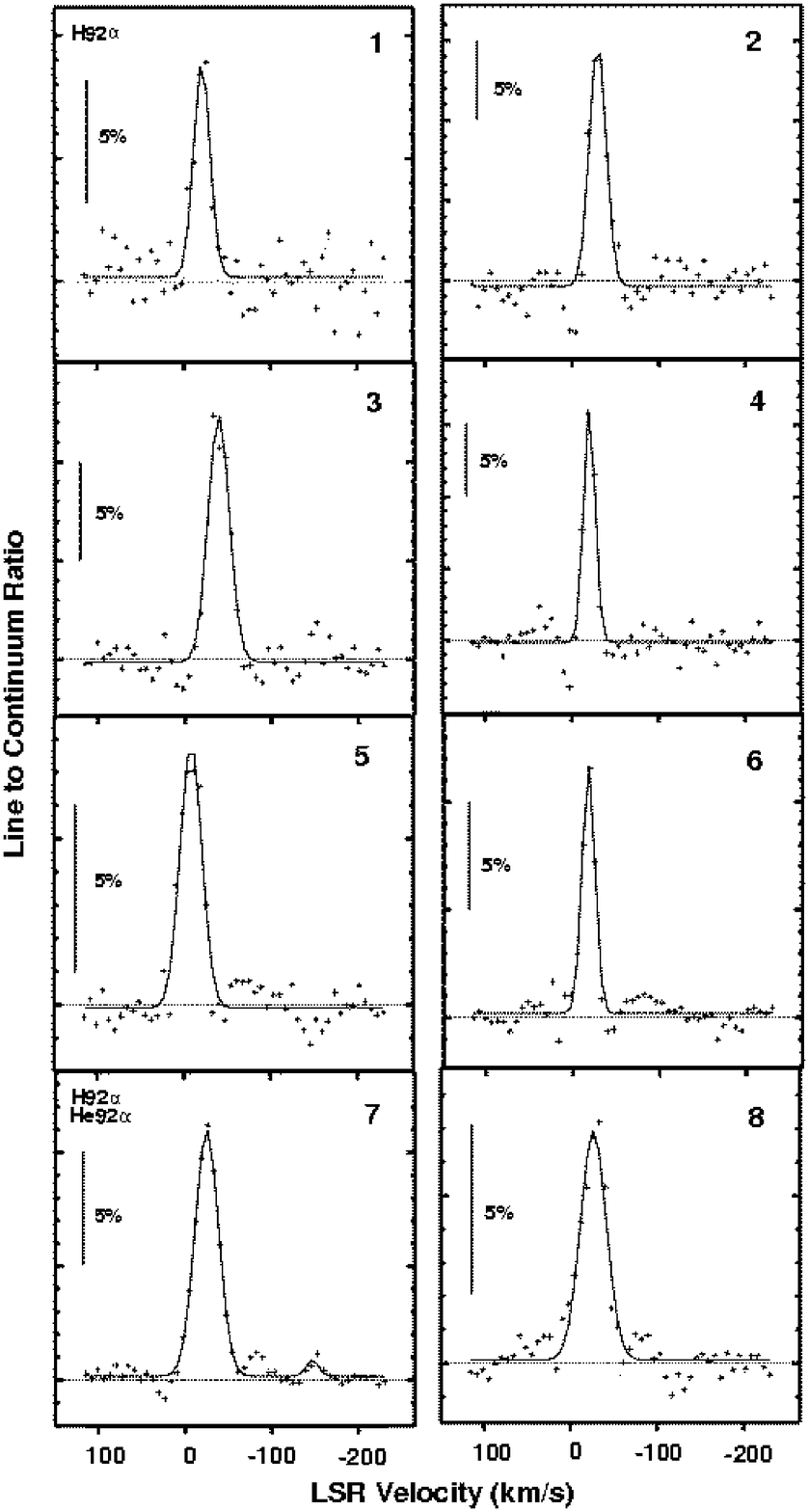}
\caption{The spatially integrated, continuum-weighted H92$\alpha$ line
profiles for the 25 regions in the Arched Filaments shown in Figure
7. The format is identical to Figure 6. In most of the regions, a
single Gaussian profile was fit to the data. Profiles 12 and
14 show a double component profile with two components. He92$\alpha$ lines were fit at the 2-3 $\sigma$ level in Profiles
7, 9, 16 \& 19. Properties of the Gaussian fits are listed in Table 7.}
\end{figure}

\begin{figure}[t!]
\plotone{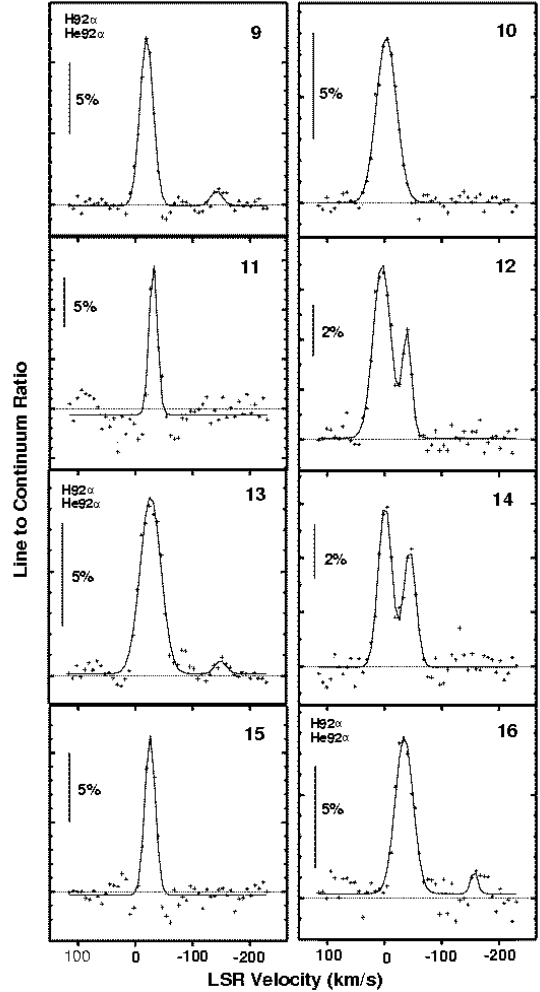}
\caption{H92$\alpha$ profiles 9-16 as shown in Figure 8.}
\end{figure}

\begin{figure}[t!]
\plotone{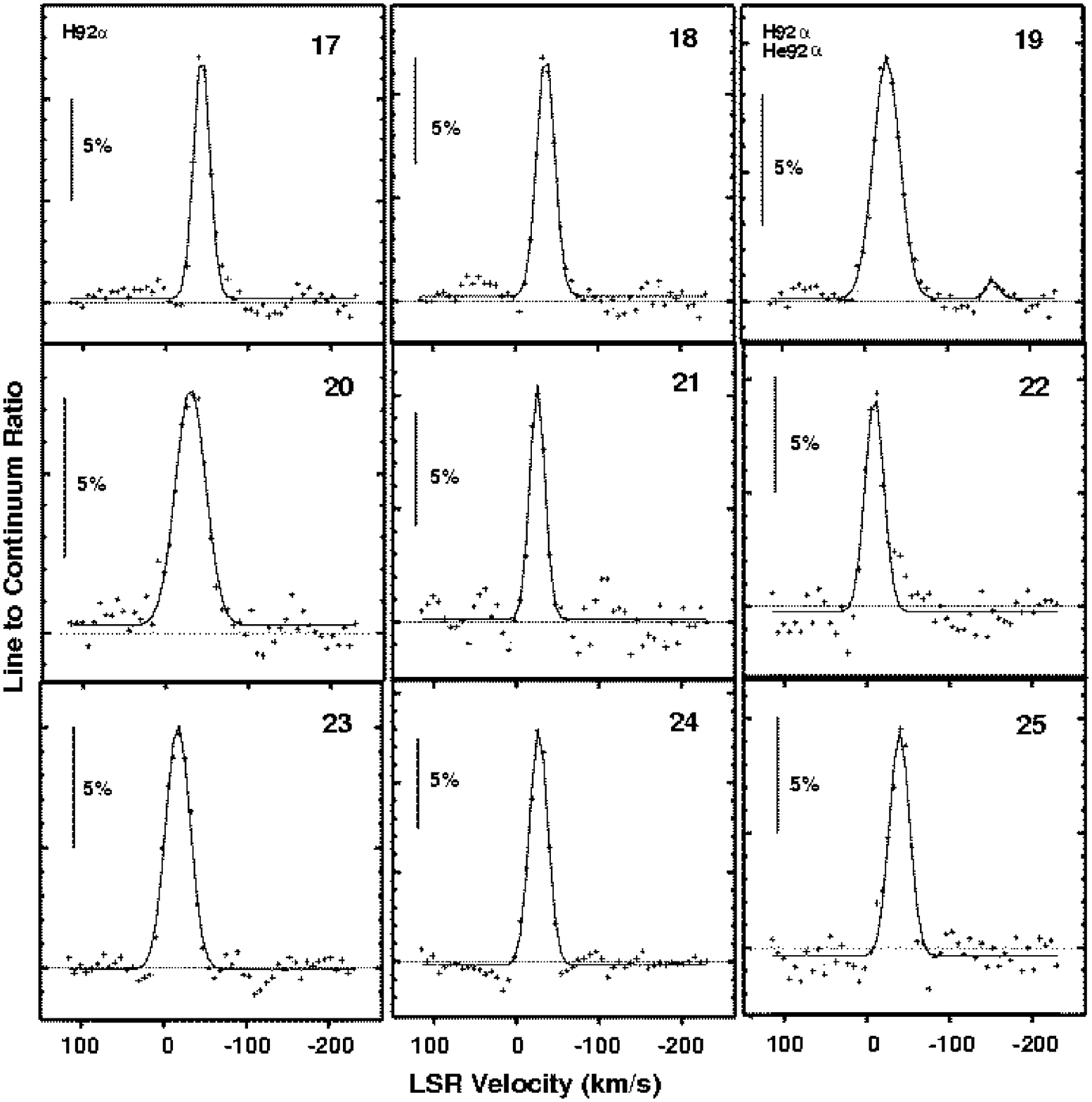}
\caption{H92$\alpha$ profiles 17-25 as shown in Figure 8.}
\end{figure}

\subsection{Physical Conditions of the Ionized Gas}

The line-to-continuum ratios across the Arched Filaments range from
0.04 to 0.17, with an average value of 0.11 (Table
7), consistent with a typical line-to-continuum ratio in the H92$\alpha$
line of 0.1 (calculated by assuming typical LTE conditions for
Galactic center H II regions: $\Delta$V\ab30 \kms~and
T$_e$$^*$\ab6000 K). The FWHM line widths in the Arched Filaments range from 15 to 44 \kms,
with an average value of 28 \kms, similar to the radio recombination line widths observed for other Galactic center H II regions
($\Delta$V\ab 27 \kms~in a survey by Downes et al. 1980;
$\Delta$V\ab33 \kms~in SgrB1 by Mehringer et al. (1992); and
$\Delta$V\ab35 \kms~in the Sickle and Pistol by Lang et
al. (1997)). Typically, the line widths in Table 7 with values $>$ 35
\kms~(corresponding to Profiles 10, 13, 19, \& 20) are due to the
substantial velocity gradients present over the integration
regions. Figure 12 also illustrates that over most of the Arched
Filaments, the line widths are 20$-$30 \kms. In a few places, the line
widths exceed 50 \kms. These positions are the sites of double
profiles, and therefore the broad line widths are likely due to a poorer
fit by a single Gaussian component. 

\begin{figure}[t!]
\plotone{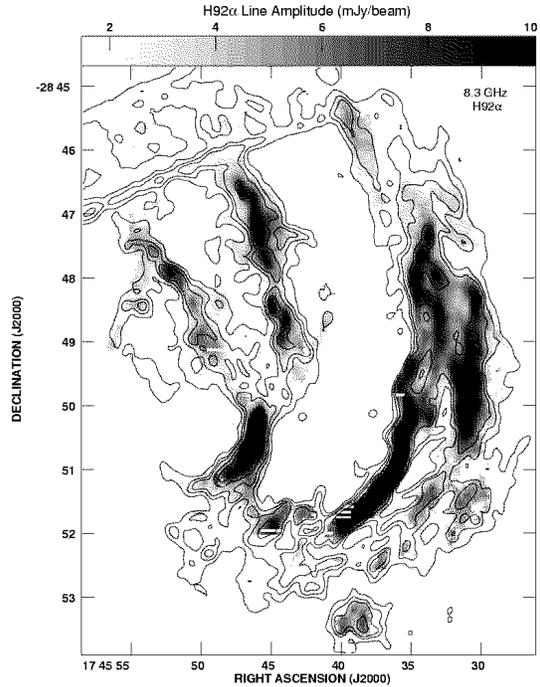}
\caption{Distribution of H92$\alpha$ line amplitude based
on the pixel-by-pixel Gaussian fits to the data. The greyscale
represents the range (a) 1 to 10 mJy \beam, (b) 15 to 50 \kms, and
(c) 4000-10000 K,  where the resolution is 12\farcs8 $\times$ 8\farcs1
in the line images. The contours in all three panels represent 8.3 GHz primary
beam-corrected continuum emission at levels of 6, 13, 18, 36, 60, and 120  mJy \beam, with a resolution of 7\farcs78 $\times$ 6\farcs61, PA=$-$1.0.}
\end{figure}

\begin{figure}[t!]
\plotone{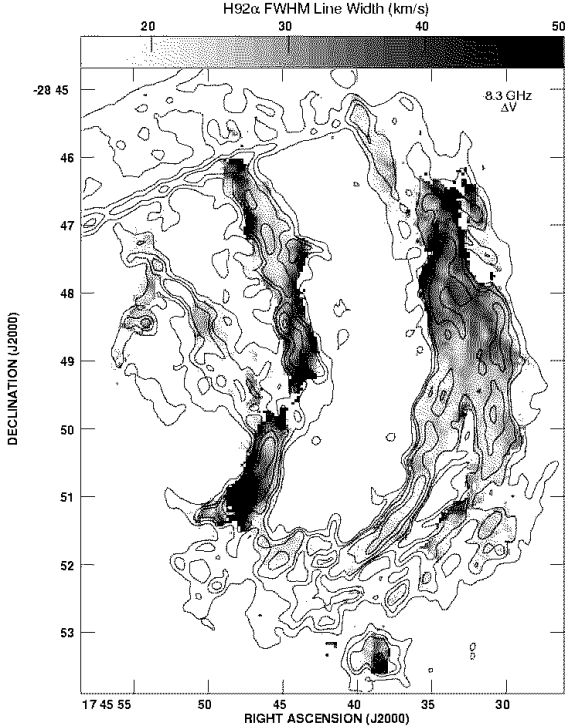}
\caption{Distribution of H92$\alpha$ FWHM line width as shown in Figure 11.}
\end{figure}

\begin{figure}[t!]
\plotone{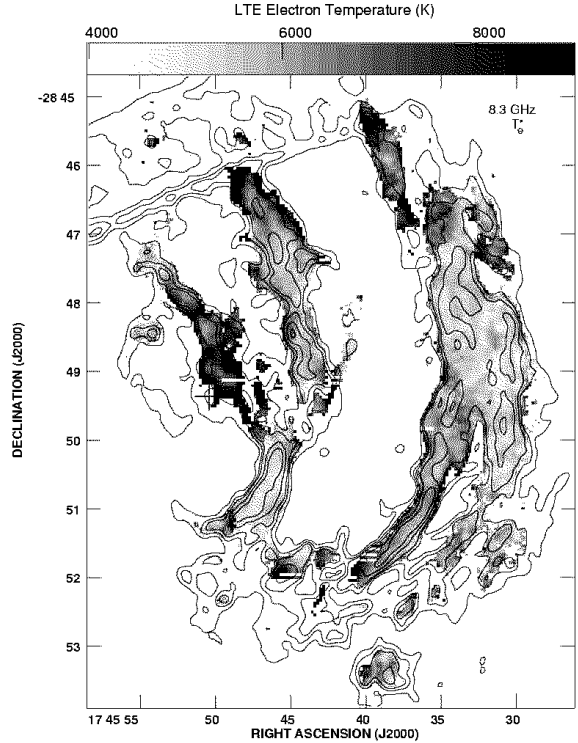}
\caption{Distribution of LTE electron temperature as shown in Figure 12.}
\end{figure}

\subsubsection{Electron Temperatures}
LTE electron temperatures, T$_e^*$, are calculated for the
25 regions, based on the measured values of T$_l$/T$_c$, $\Delta$V,
and an assumed value of Y$^+$=0.06 (eq. [22]; Roelfsema \& Goss 1992). Although the He92$\alpha$ line was only detected in some regions of the
source, this average value was used for all the regions, since
\T~depends only weakly on Y$^+$; assuming a value of Y$^+$=~0 only
increases \T~by a few percent. 
To determine the importance of non-LTE effects, the LTE departure
coefficients (b$_n$,$\beta$$_n$) are calculated based on average values of the
electron density (n$_e$\ab~300 cm$^{-3}$) and continuum optical depth
($\tau$$_c$\ab0.0015) (see Table 4). The departure coefficients are used to
derive non-LTE temperatures for several regions (eq. [23] of
Roelfsema \& Goss 1992).  At most, the non-LTE temperatures are
decreased by 2\% from the LTE values. Therefore, the non-LTE
corrections can be considered to be a negligible effect in the Arched Filaments, and we can assume that the emission occurs
under LTE conditions. In addition, Shaver (1980) points out that for a
given emission measure (EM), there is an observing frequency where the
non-LTE effects can be considered negligible ($\nu$ = 0.081
EM$^{0.36}$ (GHz)). Following Shaver  (1980), for the H92$\alpha$ line at 8.3 GHz,
the EM corresponding to LTE conditions has a value of 3.8\x10$^5$ pc
cm$^{-6}$. The values of EM listed in Table 4 are in the range of 1.2 to 3.0\x10$^5$
pc cm$^{-6}$, compatible with the assumption that the H92\al~lines in
the Arched Filaments are emitted under LTE conditions.  

The values of T$_e^*$ in the Arched Filaments range from 5000\p300 to 8900\p900 K,
with an average value for the LTE electron temperature of 6200 K. Similar T$_e^*$
have been measured in other H II regions in the Galactic center: 7000 K in SgrA West (Roberts et
al.  1993), 6400 K in the ``H'' regions (Zhao et al. 1993), and 5500 K in
the Sickle (Lang et al. 1997). 

Several of the regions in the Arched Filaments (Profiles 4, 6, 11, \&
12) show very narrow FWHM line
widths ($\Delta$V$<$20 \kms) and represent some of the narrowest
lines observed in Galactic center H II regions. Such narrow lines can
place upper limits on the electron temperatures of the ionized
gas. The Doppler temperature, T$_D$, is defined by the line width
for thermal
motion in the absence of turbulence and pressure broadening (T$_D$=21.8~($\Delta$V)$^2$ K, where $\Delta$V is in \kms). 
Narrow line widths have been observed in only a small number of sources and
provide an important demonstration that electron temperatures
as low as 4000$-$5000 K do exist in some nebulae (Shaver et al. 1979,
1983; Kantharia et al. 1998).  
Values of T$_D$ in regions where the lines are very narrow can be compared with \T~to check for consistency. For this comparison, values for T$_D$ in the 25 regions are listed in Table
7. The narrowest lines in the Arched Filaments (15.8, 16.7, and 17.5
\kms) place upper limits on the electron temperatures in these regions
of 5400\p400, 6600\p400, and 6000\p600 K respectively. 
In most of the regions of the Arched Filaments, the measured electron
temperatures are consistent with the Doppler temperatures within the
errors. In two regions (4 and 6) there is a discrepency (only two
sigma) between the Doppler temperature and the measured \T.

Figure 13 shows the distribution of \T~across the Arched
Filaments. Over most of the source, the \T~are \ab6000 K, but along
the northern edges of E2 and W1, and along the middle of the E1
filament (Regions 4, 5, 6, and 7), the \T~appear to increase up to 10,000 K.
A likely explanation for the increased \T~in the northern portion of
E2 and W1 is that the continuum emission in these regions is contaminated by the non-thermal contribution of the NTFs in the Radio Arc; the
line-to-continuum ratio is therefore underestimated, and T$_e$ is therefore overestimated, since
T$_e$ $\propto$ (T$_l$/T$_c$)$^{-0.87}$.
By measuring the continuum emission in a region of the Arched Filaments adjacent to the Radio
Arc NTFs, we estimate that 40\% of the continuum at the positions of the NTFs is nonthermal, and we correct the values of T$_l$/T$_c$ and
\T~accordingly. Similar corrections were made to the
line-to-continuum ratios in the Sickle H II region, where the values
were significantly reduced due to the non-thermal contribution from
the NTFs in the Radio Arc, which intersect the H II region at several positions (Lang et al. 1997).  

\subsection{Velocity Field}
Figure 14 shows the distribution of central velocities in the Arched
Filaments, which range from $-$70 to +15 \kms. The most
impressive feature of Figure 14 is the presence of remarkable
velocity gradients along the extent of each of the Arched Filaments.  
Figures 15-17 show position-velocity diagrams for each filament (W1, W2, E1, E2, and G0.10+0.02). These diagrams were created to illustrate velocity as a function of position following closely the ridges of emission in each case. The velocities were sampled starting at the northernmost point (see captions for Figures 15-17) and continuing southward along each filament. 
The sense of the velocity gradient in W1 and W2 can be characterized by
increasingly negative velocities southward along both filaments.
Figures 15 a and b show that the velocities steadily decrease from
$-$10 \kms~to $-$60 \kms~in both cases.
The most negative H92$\alpha$ line emission in the Arched Filaments (V
$<$ $-$50 \kms) occurs at the southern extents of W1 and W2, in the
vicinity of \ad=17 45 36.0, $-$28 51 30.     

\begin{figure}[t!]
\plotone{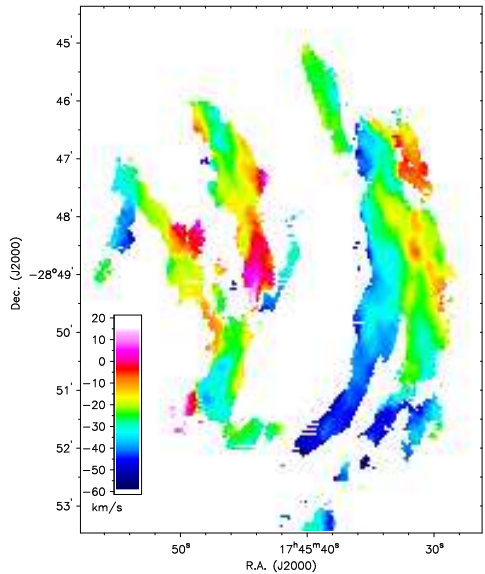}
\caption{Distribution of LSR central velocity shown in false color
representing velocities $-$60 to +20 \kms, based on Gaussian fits to each pixel above a 4$\sigma$ level.}
\end{figure}

\begin{figure}[t!]
\plotone{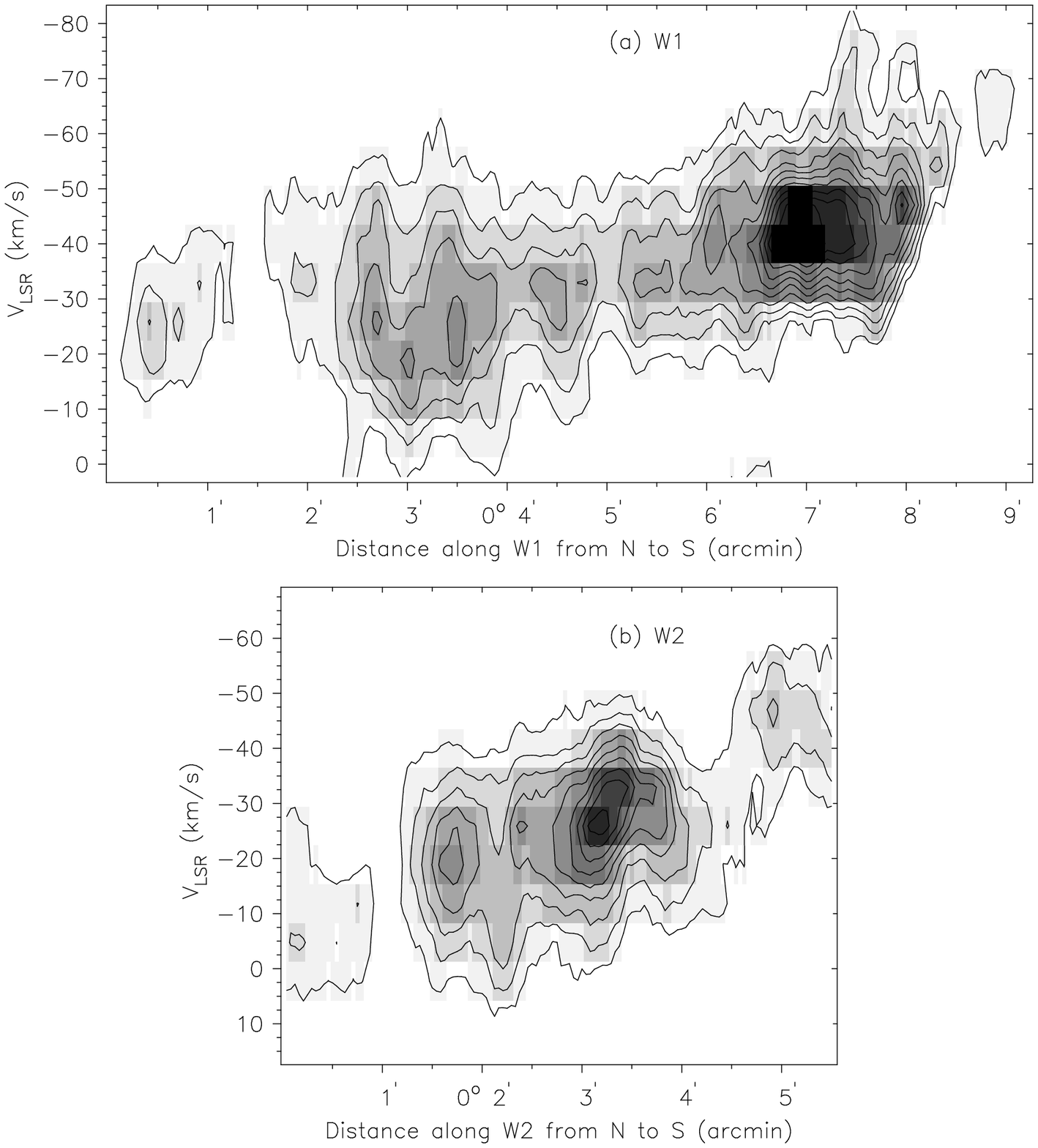}
\caption{Distribution of velocity as a function of position from N to S, closely following the ridges of emission in each filament: (a) W1 (starting at \ad=17 45 43.0, $-$28 44 30), (b) W2 (\ad=17 45 32.0, $-$28 46 00),  (e) G0.10+0.02 (\ad=17 45 45.0, $-$28 50 00). Velocity was sampled in a direction perpendicular to the filament for (f) E1 (at $\delta$\ab$-$28 48 30), and (g) E2 (at $\delta$\ab$-$28 47 30). The contours in all figures represent the H92$\alpha$ line intensity between 2 and 20 mJy \beam~at 2 mJy \beam~levels.}
\end{figure}

\begin{figure}[t!]
\plotone{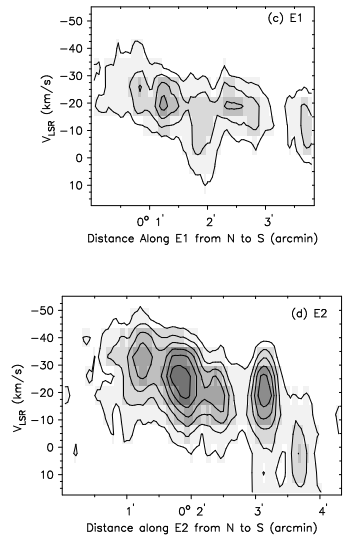}
\caption{Distribution of velocity as a function of position from N to S, closely following the ridges of emission in each filament: (c) E1 (\ad=17 45 55.0, $-$28 47 00), (d) E2 (at \ad=17 45 50.0, $-$28 46 00). The contours in all figures represent the H92$\alpha$ line intensity between 2 and 20 mJy \beam~at 2 mJy \beam~levels.}
\end{figure}

\begin{figure}[t!]
\plotone{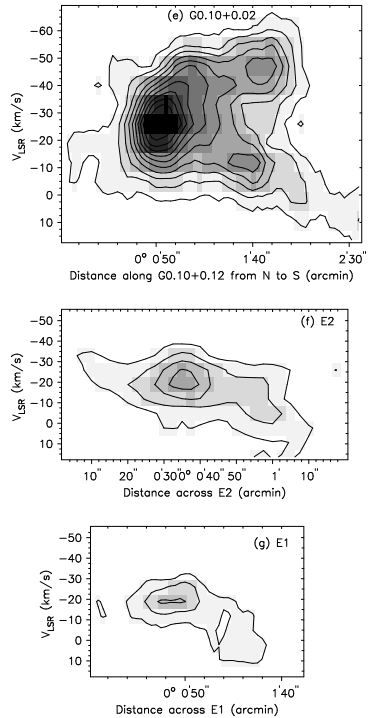}
\caption{Distribution of velocity as a function of position from N to S, closely following the ridges of emission in each filament: (e) G0.10+0.02 (\ad=17 45 45.0, $-$28 50 00). Velocity was sampled in a direction perpendicular to the filament for (f) E1 (at $\delta$\ab$-$28 48 30), and (g) E2 (at $\delta$\ab$-$28 47 30). The contours in all figures represent the H92$\alpha$ line intensity between 2 and 20 mJy \beam~at 2 mJy \beam~levels.}
\end{figure}

\begin{figure}[t!]
\plotone{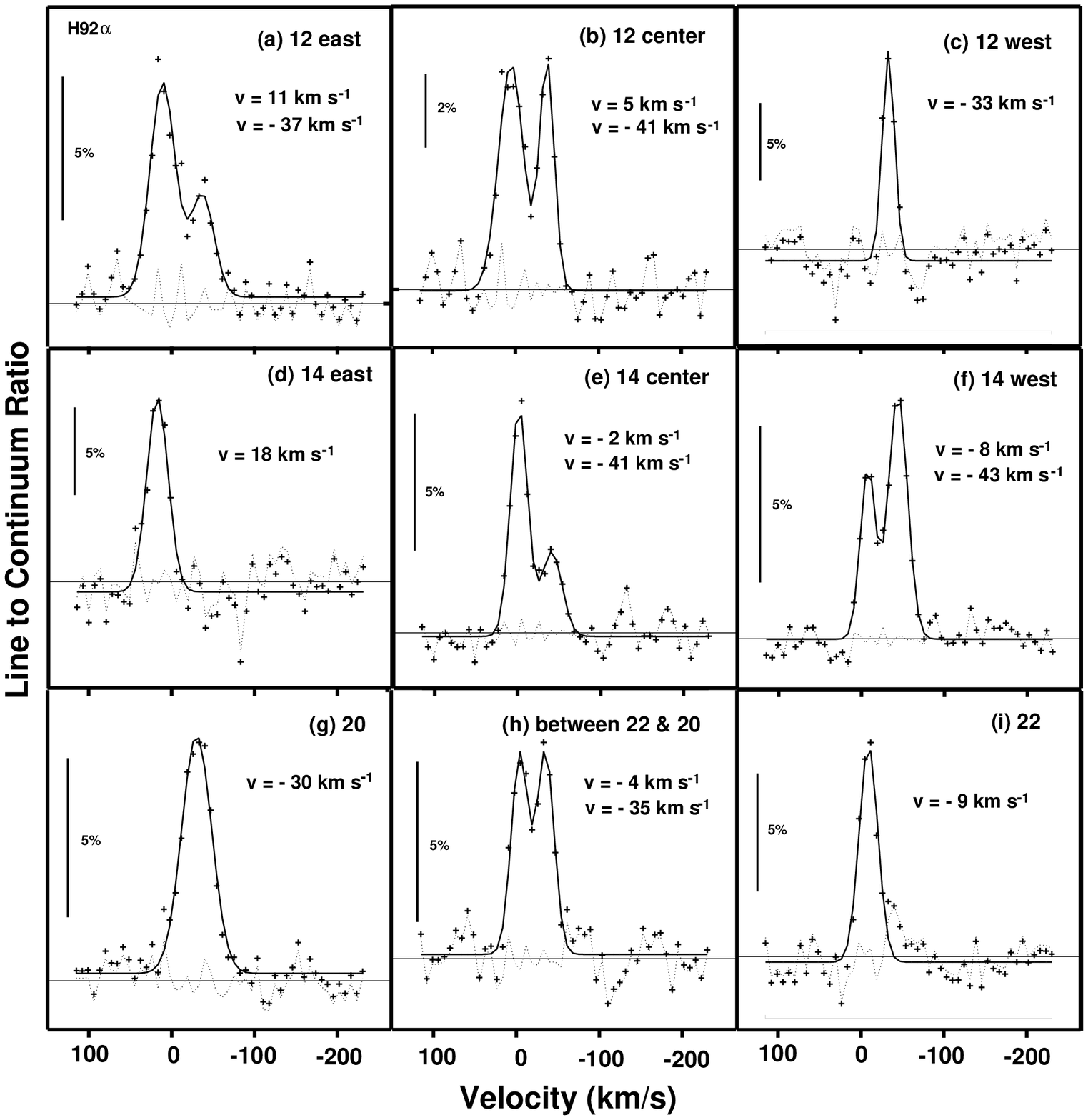}
\caption{Three series of H92$\alpha$ profiles from each of the three regions which exhibit double peaked profiles. From left to right, the panels represent continuum-weighted integrated profiles sampled at 15\arcsec~intervals from East to West within each integration region. The labelling is identical to the profiles shown in Figures 6 and 8-10.}
\end{figure}

The velocity structure in E1, E2, and G0.10+0.02 is more complex, and
the velocity gradients do not have the same sense, or the same
degree of continuity, as
those in the W filaments. In the northernmost region of E1, a small spur of line emission which is
oriented nearly perpendicular to the eastern edge of the filament
(at \ad=17 45 54.3, $-$28 48 10) has velocities which are more
negative than in the rest of the filament (V\ab$-$40 \kms~in the spur
compared with $-$20 \kms~across E1). The velocity in this spur
decreases southward along its length, similar to the sense of the
gradient in W1 and W2.  In the middle of E1 the velocities can be
characterized with values of $-$15 to 0 \kms, which become only slightly more negative
(V\ab$-$20 \kms) in some regions to the south. Over parts of E1, the
gradient has the same sense as W1 and W2, but most of the
ionized gas has velocities which are less negative than those in the W
filaments. 

The sense of the velocity gradient in E2 is nearly opposite to that in
the W filaments; the velocities become more
positive southward along E2 (Figure 16d).
In fact, there are several concentrations of {\it positive} velocity
emission in E1 and E2, which are not present in the W filaments. The
most extreme positive velocities in the Arched Filaments
(V\ab+10 \kms), occur along the southern edge of E2.  
At this position (\ad=17 45 42.6, $-$28 49 30), there is also a spur of negative velocity
emission located on the west side of E2 and extends over \ab2\arcmin~(5 pc). The
velocity gradient along this component has the same sense as in the W1
and W2 filaments. 

G0.10+0.02 has velocities which range from about $-$20 \kms~near its
northern portion, to values which are increasingly more negative ($-$40
\kms) along its southern extent, also resembling the gradients in W1 and W2. In addition, G0.10+0.02 has an unusual spur of positive velocity emission 
located at its southeastern edge.

The magnitudes of the velocity gradients vary across the Arched
Filaments and represent some of the most
extreme and coherent gradients in the Galaxy. In W1, the velocity smoothly decreases from N to S, equivalent to a
change of 40 \kms~over 7\arcmin, or 2.3 \kms~\pc. Along W2, a
gradient of \ab5 \kms~\pc~is present in the N-S direction. In
E2, the velocity ranges from $-$30 \kms~to +20 \kms, corresponding to a
velocity gradient of 7 \kms~\pc, whereas in the E1 filament,
the velocity varies by 2.4 \kms~\pc~in the N-S direction.
The velocity gradients with the largest magnitude in E1 and E2 occur in
a direction {\it perpendicular} to the long axis of the filament. 
Figures 17f and g show position-velocity diagrams for slices of
1\arcmin$-$1\farcm5~(2.5$-$3.8 pc) length taken across E1 and E2 in a direction
nearly perpendicular to the long axis of the filament where it is
apparent that the velocities are changing rapidly. In both
cases the gradients in this direction are \ab16 \kms~\pc~, several
times larger than the gradients observed in the N-S direction in the
other filaments. The velocity gradients in the Arched Filaments are
only surpassed in magnitude by the nearby ionized streamers of SgrA
West, which surround the nuclear black hole, SgrA$^*$, and have
velocities varying from $-$200 \kms~to 200 \kms~over \ab3 pc (Roberts
\& Goss 1993).

\subsubsection{Double-Peaked H92$\alpha$~Profiles}

Three of the profiles have double-peaked structure: Regions 12 and 14, as well as a a double profile between Regions 20 and 22 (at \ad=17 45 33.5, $-$28 47 30). At each of these positions there are large differences in the velocities of adjacent filaments as can be seen in Figure 14. The double profiles in the Arched Filaments do not have a systematic or symmetric distribution across any part of the source, such as might be indicative of an expanding, H II shell. Therefore, it appears that the double profiles simply occur at the interfaces of two gas components which have different velocities. These velocities, however, are consistent with the previous velocity results presented above.

Figure 18 illustrates that a simple superposition of ionized gas components which have largely different velocities can explain the double profiles in all three regions. Figure 18 shows three integrated profiles from each of these double-peaked regions; each profile has been sampled at \ab15\arcsec~intervals from east to west across the region (corresponding to left to right in Figure 12).
The central profile in Region 12 (Figure 18 b) shows a double profile with velocities of +5 \kms~and $-$41 \kms. In the east and west parts of Region 12 (Figures 18 a and 18 c), the dominant components of velocity are +11 \kms~and $-$33 \kms, respectively, which appear to correspond to the two velocity components apparent in Figure 18. The positive velocities likely arise from the ionized gas in the adjacent part of E2, which is characterized by a similar velocity range (+5 to $-$5 \kms). The negative velocity emission arises from a negative velocity spur located to the south and east of E2 (Region 11) which was previously discussed and found to have velocities in the range of $-$30 to $-$45 \kms. 
A similar pattern is detected in the double components of Regions 14 and between W1 and W2. The eastern portion of Region 14 (Figure 18d) is characterized by a positive velocity (V=18 \kms), in contrast to the negative velocity ($-$40 \kms) in the western part of this region (Figure 18e). At this position, the superposition of the negative-velocity emission in G0.10+0.012 and a positive-velocity component produces the two peaks in Profile 14. The position-velocity diagram of G0.10+0.02 (Figure 17e) also shows the two velocity components which differ by \ab60 \kms. There is a large velocity gradient in this positive-velocity component and the velocities range from +20 to $-$10 \kms. The two peaks in Figure 18 h have velocities of $-$4 and $-$35 \kms, corresponding closely to the central velocities of the W1 and W2 filaments (Figures 18g and 18i). {\it The presence of such double-peaked profiles illustrates that this H II complex consists of a series of independent filaments of ionized gas having different kinematics and different displacements along the line of sight.}

\begin{figure}[t!]
\plotone{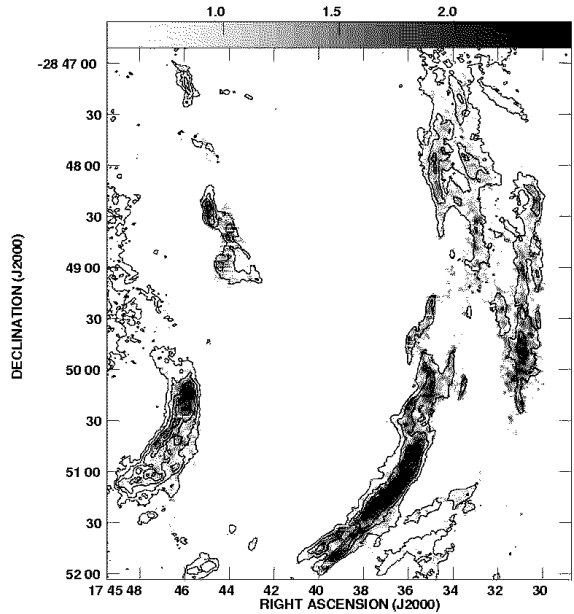}
\caption{High resolution (3\farcs7 $\times$ 2\farcs7, PA=44\fdg6) H92$\alpha$
line amplitude shown in greyscale, superposed by the high-resolution
(2\farcs3 $\times$ 1\farcs6, PA=64\fdg2) primary beam corrected 8.3 GHz continuum emission of the Arches 1 region, centered at \ad=17 45 35.0, $-$28 50 00. The greyscale represents the range of 0 to 2.5 mJy \beam, and the contour levels represent 1.5, 2, 3, 4, 5, 6, 7, 8 mJy \beam.}
\end{figure}

\begin{figure}[t!]
\plotone{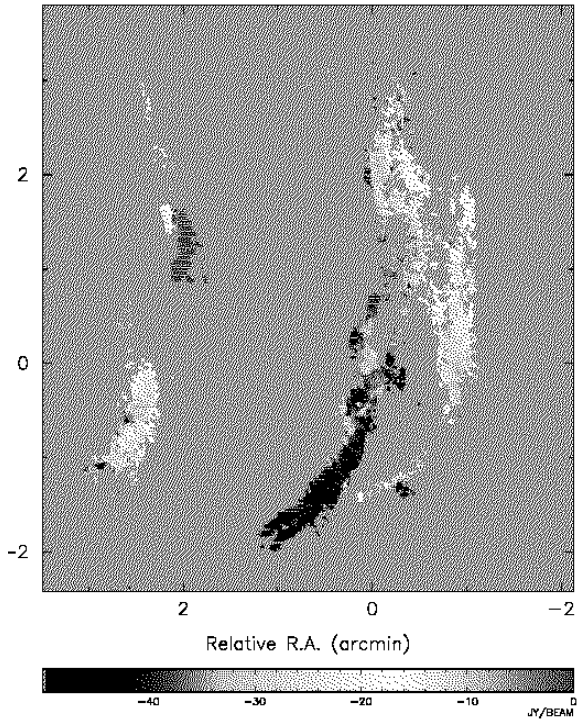}
\caption{Distribution of LSR velocity shown in greyscale, based on
fits to each pixel above a 4$\sigma$ level in Arches 1 field. The
resolution of the line emission is 3\farcs7 $\times$ 2\farcs7,
PA=64\fdg2. The greyscale represents 0 to $-$50 \kms.}
\end{figure}

\subsection{High Resolution H92\al~Line Observations}
The CnB and DnC array observations of a single field in the
southwestern portion of the 
Arched Filaments were combined to produce a higher resolution (3\farcs6 $\times$ 2\farcs7, PA=45\arcdeg) H92$\alpha$ line cube. 
This region was chosen since it has strong emission and contains the
intersection of the Northern Thread NTF with both the W1 and W2
filaments. Figure 19 shows the distribution of H92$\alpha$
emission in this region in greyscale, with 8.3 GHz continuum contours. The strongest emission arises from along
the W1 filament, in a series of ``knots'', with most of the emission
concentrated at G0.07+0.04. The emission along the rest of W1 is tenuous, and along portions of W1, the line emission is
dramatically edge-brightened, so that the center of the filament appears 
hollow. The velocity field of the high resolution data (Figure 20) resembles
the overall structure of the lower resolution image (Figure 14). The
velocity gradients in the W1 and W2 filaments are consistent with those
measured in the lower resolution data: 2 to 4 \kms~pc$^{-1}$. 

At the locations of intersection between W1, W2 and the Northern
Thread NTF, there do not appear to be any significant changes or
discontinuities in the line properties (FWHM line width or velocity
field). If there were significant discontinuities, this would indicate that a physical interaction is taking place between the
ambient magnetic field and the ionized gas. However, the
velocity along the southern portion of W1 varies as smoothly as in the lower
resolution data (Figure 14), becoming increasingly negative toward the
southern extent of the filament. In fact, the central velocity changes more
abruptly in the northern portion of W2, where values range from \ab0
\kms~to $-$35 \kms~over several pc. 
Thus, large disturbances in the velocity field at the positions of
projected intersection with the Northern Thread NTF are not observed.   

\section{Discussion}
\subsection{Morphology}
The close correspondences between the
velocities and morphology of the ionized and molecular gas in the
Arched Filaments that Serabyn \& \gusten~(1987) illustrated indicate that the Arched Filaments represent the ionized edge of 
the $-$30 \kms~cloud. The Arched Filaments therefore derive their morphology in part from the
distribution of molecular gas. 
Tidal disruption due to the large differential gravitational forces near the Galactic center is likely to have influenced the morphology of the $-$30 \kms~cloud. In fact, Serabyn \& \gusten~(1987) propose that this molecular cloud is just at the stability limit for tidal disruption (n\ab5 $\times$10$^4$ cm$^{-3}$ for a radius of \ab30 pc) and has become tidallly unstable as it has fallen in toward the Galactic center from an outer radius. 
The elongated morphology of the Arched Filaments therefore can be ascribed in part to such tidal shearing of the cloud, given that the overall length of
the filaments (\ab20 pc) is comparable to their projected radial displacement
from the Galactic center (\ab25 pc).  The CS emission in the inner 0.5-1\arcdeg~from the survey of Bally et al. (1988) shows that the clouds  
 in this region are elongated along the Galactic plane, whereas at
1\fdg5, the clouds are oriented orthogonal to the plane. 

Other evidence for shearing of the cloud includes the orientation of the magnetic field in the Arched Filaments, traced by
the far-IR polarization vectors (Morris et al. 1992; Morris et al. 1995). The inferred magnetic field
orientation is aligned along the length of the
filaments, suggesting that shearing occurs in a direction parallel to
the filament long axis. This conclusion can be drawn for any arbitrary
initial magnetic field geometry as long as the kinetic energy density
in the shear motions exceeds the energy density in the magnetic field,
in which case the magnetic field will be deformed until it is oriented in a
direction parallel to the shearing (Morris et al. 1992).
Of course, the location of the ionizing source in relation to the molecular cloud will also affect the morphology of the Arched Filaments. 

Several clues suggest that the
ionized gas in the Arched Filaments is likely to be on the near edge
of the molecular cloud.  Observations of the Brackett-$\gamma$ line at
2.166 $\mu$m reveal an extinction toward the Arched
Filaments consistent with the general extinction toward the Galactic
center, indicating that the ionized gas is on the near side of the
molecular cloud (Figer 1995; Cotera et al. 2000). These results are consistent with the fact that the
observed H92\al~velocities are blueshifted at some locations relative
to the velocities of the molecular gas at the same positions (Serabyn
\& \gusten~1987). A further, more detailed study of the relative
locations and velocities of ionized and molecular gas in the Arched
Filaments is the subject of Paper II (Lang et al., in prep.). 

\subsection{Kinematics}

The H92$\alpha$ line observations provide several important clues for
understanding the kinematics and velocity field of the ionized edge of
this peculiar molecular cloud. From these data we have deduced the
sense and the magnitude of the velocity gradients at several positions
across the source. The velocity gradients along the W1 and W2 Arched Filaments
are well-organized and remain coherent over extents as large as 20 pc.  
It therefore seems appropriate to describe the velocity field of 
the Arched Filaments in terms of the orbital motion of the larger,
underlying molecular cloud about the Galactic center. 

\begin{figure}[t!]
\plotone{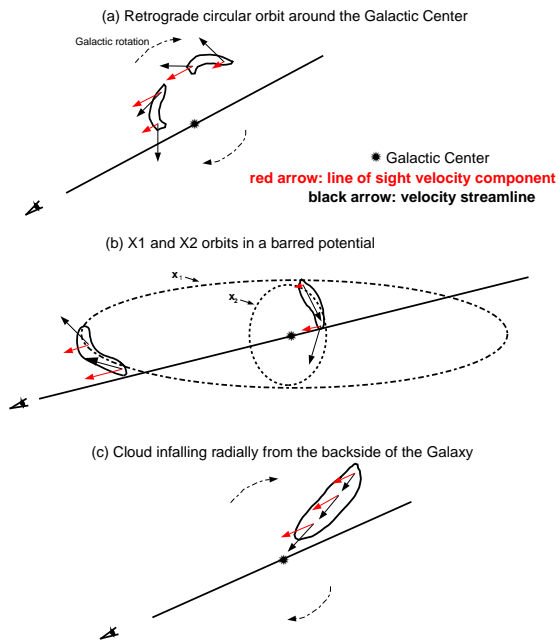}
\caption{Schematic diagram of possible orbits around the Galactic center of the $-$30 \kms~molecular cloud underlying the Arched Filaments.}
\end{figure}

\subsubsection{Possible Orbits}
{\bf Retrograde Circular Orbit}. A simple possibility is to place
    a tidally-stretched molecular cloud on a circular orbit with a direction
    of motion counter to Galactic rotation, consistent with the negative
    velocities of both the ionized and molecular gas. The origin of
    such a retrograde cloud is unknown, but Serabyn \& \gusten~(1987) propose
    that it has infallen from a large radius and has been streched in the process. In the presence of the large gravitational potential gradient near the Galactic center (where, at 30 pc, the total enclosed mass is
    \ab2 \x~10$^{8}$ M$_{\sun}$; Genzel \& Townes 1987), a cloud will be
    sheared in a direction perpendicular to the gradient in the potential and will become elongated in that direction. The schematic shown
    in Figure 21a illustrates two possible locations of this cloud on
    a counter-rotating orbit about the center of the Galaxy. As a result of shearing, the leading edge of the cloud is closest to the
    center of the Galaxy. The black arrows represent the velocity
    streamline of the molecular cloud, whereas the red arrows
    represent the observable, line-of-sight velocity component.  The measured velocities along most of the Arched Filaments are increasingly
    negative in a direction along the filament toward the Galactic
    center. As shown in Figure 21a, the sense of the velocity gradient
    resulting from a cloud on such an orbit is exactly opposite to the
    observed H92\al~line gradients. Therefore, the kinematics of the $-$30 \kms~cloud cannot be explained by retrograde circular motion. 

{\bf Molecular Cloud Orbits in a Central Barred Potential}. Observations of the distribution of stellar and gaseous material in the central kiloparsec
of the Galaxy indicate the presence of a stellar bar with a co-rotation radius of about 2.4 kpc,  and an estimated orientation of 15$-$20\arcdeg~to the line of
sight (Morris \& Serabyn 1996, and references therein).
Such a non-axisymmetric component in the Galaxy's potential will lead
to clouds following elongated, and therefore non-circular, orbits about the
Galactic center (Binney et al. 1991). 
These authors propose that clouds in the presence of such a potential
will first shear, and then move onto a set of closed orbits, slowly
migrating to orbits of lower energy as they lose angular momentum. The two 
families of orbits include those inbetween the co-rotation radius and the inner
Lindblad resonance, known as the ``x$_1$'' orbits (on which gas
approaches the center of the barred potential) and those which are
nested deeper in the potential well, known as the ``x$_2$'' orbits
(Contopolous \& Mertzanides 1977). Figure 21b shows a schematic of the x$_1$ and x$_2$ orbits with a molecular cloud placed on each so that the sign of the Galactic longitude, the sign of the velocity, 
and the sense of the gradient in the cloud are consistent with the
H92\al~observations. The cloud on the x$_2$ orbit is located on the
far side of the Galactic center and has a larger negative velocity component
arising from the portion closer to the Galactic center than the trailing edge.
The cloud on the x$_1$ orbit is positioned at the near extremity of its
orbit along the line of sight, so that the leading edge is leaving our sight line, producing the sense of the velocity
gradient that is observed. Both orbits can explain the so-called
``forbidden'' velocities of the $-$30 \kms~molecular cloud, but it is
necessary to further distinguish between them. 

{\bf Radially Infalling Molecular Cloud}.
Yet another possibility is that the molecular cloud is infalling radially from a larger radius behind the Galactic center and is thereby stretched in the radial direction. Figure 21c shows a sketch of
this possibility. For this scenario, the sign of the velocity and the sense of
the velocity gradient are consistent with the H92$\alpha$ data in three of the four Arched Filaments (E1, W1, W2). However, the sense of the velocity gradient in E2 is exactly opposite to these (that is, the velocities increase southward along the filament).  

\subsubsection{Comparison with Observed Kinematics}

In order to determine whether the velocity field in the Arched
Filaments can be ascribed to the x$_1$ or x$_2$ family of orbits, or to a 
radially infalling molecular cloud, one can compare the
magnitude of the predicted velocity gradient for these scenarios to that measured from the H92$\alpha$ data.

In the case of the x$_1$ and x$_2$ orbits, it is possible to compare
the H92$\alpha$ velocity gradients with the values from the simulated position-velocity ($\ell$,v) diagrams of Jenkins \& Binney (1994), 
which represent gas flows in the central Galaxy.
These authors have supplemented the orbital models of Binney et al. (1991) by carrying out sticky-particle simulations of
the molecular clouds on the x$_1$ and x$_2$ orbits, in order to
include the hydrodynamical effects of the collisions of such clouds in
the inner Galaxy and the subsequent exchange of mass and
momentum. Their simulated ($\ell$,v) plots reproduce some, but not all aspects of the $^{13}$CO parallelogram from the Bell Labs
survey (Bally et al. 1988). The gradients measured in H92\al~along W1 and W2 have magnitudes of 2$-$7 \kms~pc$^{-1}$, which have
increasingly negative velocities for decreasing longitude on the
positive longitude side of the Galaxy,  can be compared with
Figure 12 of Jenkins \& Binney (1994), which shows the resulting ($\ell$,v) diagram for the end-point (near
steady state) of their simulation at a viewing angle of \ab10\arcdeg~to the
long axis of the bar. The
velocity gradients in the corresponding region of their Figure 12 are in the
range of 0.5$-$1 \kms~pc$^{-1}$ for the x$_2$ orbits, and 0.1$-$0.2
\kms~pc$^{-1}$ for the
x$_1$ orbits. The x$_2$ orbits have velocity gradients closest in
magnitude to
those measured in the H92$\alpha$ data, yet the models underestimate what we measure by
at least a factor of two.  Jenkins \& Binney (1994) suggest that their
simulated ($\ell$,v) diagram for an intermediate time (0.4 Gyr; their
Figure 6) better
reproduces several of the features in the CO parallelogram; the
x$_2$ orbits at that time have a velocity gradient of 1$-$2
\kms~pc$^{-1}$, in better agreement with what is measured, though
still somewhat low.  

In the case of the radially infalling cloud, the velocity gradient is established as a result of the radial extent of the cloud, since the infall velocity is primarily a fixed function of radius. It is possible to calculate the work done by the central gravitational
potential on a parcel of infalling gas, and thereby to estimate the resulting velocity
gradient. Using the mass distribution of Genzel \& Townes (1987),
M$_G$(r)=2 $\times$ 10$^8$ M$_{\sun}$~$\left(\frac{R_g}{30
pc}\right)^{1.2}$, the expected velocity gradient at a Galactic radius of r=25
pc due to infall is 6.4 \kms~pc$^{-1}$. This value is consistent with
the magnitudes of velocity gradients measured along the W2 and E2
filaments (5 \kms~pc$^{-1}$ \& 7 \kms~pc$^{-1}$), and is of the same order of magnitude as the velocity gradients measured in all four filaments. 
Both the x$_2$ orbit and the radially infalling cloud, however, have difficulty explaining the opposite sense of the velocity gradient in E2 without invoking a different orientation for this filament. In addition, the 16 \kms~pc$^{-1}$ gradients which are measured in E1 and E2 in a direction perpendicular to the filaments also can not be easily explained by the two scenarios without some alteration of the filament geometry. 

Although large portions of the H92$\alpha$ velocity data are well-matched with the radially infalling cloud scenario, the possibility that this cloud resides on an x$_2$ orbit should not be completely ruled out, since some of the smaller gradients are also consistent with the gradients predicted for the x$_2$ family of orbits. On the other hand, the x$_1$ orbits can probably be ruled out, since they underestimate all
measurements of the gradient by an order of magnitude or more. Both Binney et al. (1991) and Jenkins \& Binney (1994) suggest that the giant and dense Galactic center molecular clouds
(for example, the Sgr A, B and C cloud complexes) are likely to reside on the
x$_2$ orbits. In addition, the CS emission, which traces the dense populations of clouds, overlaps in the ($\ell$, v) plane very clearly with the family of x$_2$ orbits and is not consistent with the x$_1$ orbit group (Bally et al. 1988). The kinematics of the Arched Filaments are therefore best described by a cloud residing on an orbit that is intermediate in its properties between an x$_2$ orbit and pure radial infall. Determining the distance to the $-$30 \kms~cloud relative to the bulk of stars in the bulge will help to further
distinguish the properties of the cloud orbit. Using a model for the
stellar distribution in the inner Galaxy and comparing with images of
the near-infrared extinction can provide information on the distance
to the $-$30 \kms~cloud. However, currently
available extinction studies of the Arched Filaments region do not
provide adequate resolution for this determination (Schultheis et al. 1999), but a future study will address this question in detail.  

\begin{figure}[t!]
\plotone{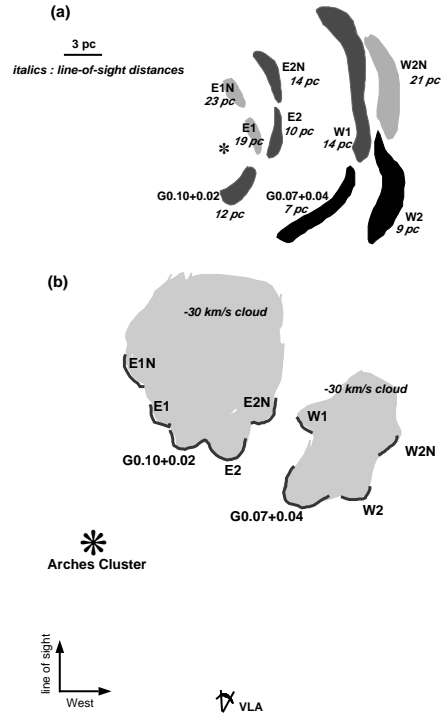}
\caption{(a) A schematic diagram showing the locations, both in projection and along the line of sight (coded in greyscale and distances in pc listed in italics) between the different portions of the Arched Filaments and the Arches cluster. (b) a simplistic diagram from a perspective above the $-$30 \kms~cloud (looking down its long axis) and showing the relative placement of the components of the Arched Filaments H II complex.}
\end{figure}

\subsection{Ionization of the Arched Filaments}

The H92$\alpha$ observations of the Arched Filaments show that the
physical conditions in the ionized gas (line-to-continuum ratios, FWHM line widths, and \T)
are similar to those found in other Galactic center H II regions, such as SgrA West
(Roberts \& Goss 1991), the ``H''-regions (Zhao et al. 1993), and the
Sickle (Lang et al. 1997), which are all photoionized by nearby massive
stars. The Arches Cluster is a stellar cluster located at the base of
the E1 filament, at \ad=17 45 50.4, $-$28 49 21.83, now thought to be
among the most massive and dense clusters
(M$_{tot}$\ab10$^4$ M$_{\sun}$, $\rho$\ab3\x10$^5$ M$_{\sun}$
pc$^{-3}$) in the Galaxy, with an age of 2 Myr (Nagata et
al. 1995; Cotera et al. 1996; Serabyn et al. 1998; Figer et al. 1999). Before the Arches cluster was well-studied, FIR and radio observations raised the question of whether it could be capable of providing the ionizing radiation necessary to account
for the ionization of the Arched
Filaments.  Based on the uniformity of the FIR fine-structure line emission
across the entire Arched Filaments region, 
Erickson et al. (1991) and Colgan et al. (1996) argue that additional stars, in excess of the G0.12+0.02 cluster as cataloged by Cotera et al. (1996), must be present and uniformly distributed, to account for the ionization of the gas. Similarly, using FIR spectroscopy, Timmermann et
al. (1996) report that the G0.10+0.02 portion of the Arched
Filaments has properties consistent with an H II region and with the
adjacent, lower excitation, photodissocation region, but that
additional stars are necessary to ionize all the filaments. Recent
studies of the Arches cluster, however, indicate that the
number of O-stars in this cluster exceeds 150 (Figer et al. 1999),
likely to provide the ionizing radiation to account for the uniformity of the FIR ionization tracers. Based on
infrared and radio observations, Morris et al. (2001) calculate the
number of ionizing photons from the Arches cluster to be 4.4
\x~10$^{51}$ photons s$^{-1}$, comparable to the estimate of Serabyn
\& \gusten~(1998) based on a rough stellar census, and at
least an order of magnitude more than the initial estimates from
Cotera et al. (1996). 
 
In order to determine whether the ionization of the entire Arched Filaments
complex can be accounted for by the Arches Cluster, we compare the
ionizing flux based on the 8.3 GHz continuum flux density
with the ionizing flux intercepted by different portions
of the Arched Filaments. As in $\S$3, calculation of the requisite number of
Lyman continuum photons (N$_{Lyc}$) is based on \T=6200 K (except for the area of E1, which has \T=8900
K) and n$_e$=300 cm$^{-3}$. We also assume that the solid angle
subtended at the Arches cluster by each portion of the Arched Filaments
can be scaled from the observed solid angle, and we assume initially that the distance between the
portions of the Arched Filaments and the cluster is the projected
distance. This latter assumption provides a lower limit to the ionizing flux; if
the Arches cluster cannot explain the ionization at the projected
distances to the filaments, then additional stars would be required.
Table 8 lists the nine regions of the Arched Filaments
(column one), the projected distance between the cluster and the
region (column two), the dimensions on the sky of each region (column
three), the number of Lyman continuum photons (N$_{Lyc}$) intercepted per second by this volume of gas, assuming that the cluster has an ionizing flux of 4.4
$\times$ 10$^{51}$ s$^{-1}$ (column four), the 3.6 cm
continuum flux density of each region (column five), and the
corresponding value of N$_{Lyc}$ (column six). 

\begin{deluxetable}{lccccccc}
\tablewidth{0pt}
\tablecaption{Ionization Parameters of the Components of the Arched Filaments}
\tablehead{
\colhead{Region}&
\colhead{Projected}& 
\colhead{Region}&
\colhead{N$_{Lyc}$}&
\colhead{S$_{\nu}$}&
\colhead{N$_{Lyc}$}&
\colhead{Total}&
\colhead{LOS}\\

\colhead{}&
\colhead{Distance}&
\colhead{Size}&
\colhead{Intercepted}&
\colhead{8.3 GHz}&
\colhead{Required}&
\colhead{Distance}&
\colhead{Distance}\\

\colhead{}&
\colhead{to Cluster}&
\colhead{(pc)}&
\colhead{(photons s$^{-1}$)}&
\colhead{(Jy)}&
\colhead{(photons s$^{-1}$)}&
\colhead{to Cluster}&
\colhead{to Cluster}}

\tablecolumns{8}
\startdata
G0.10+0.02	&4.5 pc	&4.4\x1.5	&1.1\x10$^{50}$	&2.0	&1.5\x10$^{49}$	&13	&12\\
G0.07+0.04	&8.3 pc	&5.6\x0.8	&2.2\x10$^{49}$	&1.9	&1.4\x10$^{49}$	&11	&7\\
E2S		&3.8 pc&1.0\x2.5	&4.4\x10$^{49}$	&1.0	&7.3\x10$^{48}$	&11	&10\\
W2	&10.6 pc	&5.6\x1.3	&1.5\x10$^{49}$	&1.8	&1.4\x10$^{49}$	&14	&9\\
E1N	&2.5 pc	&5.0\x1.3	&3.5\x10$^{50}$	&0.6	&4.0\x10$^{48}$	&24	&23\\
W1	&9.0 pc		&5.6\x1.3	&3.0\x10$^{49}$	&1.1	&8.7\x10$^{48}$ &17	&14\\
W2N	&10.8 pc	&5.6\x1.3	&2.2\x10$^{49}$	&0.7	&5.3\x10$^{48}$	&24	&21\\
E2N	&6.3 pc &4.5\x1.3&4.4\x10$^{49}$&1.4	&1.0\x10$^{49}$ &15	&14\\
E1	&3.0 pc	&3.8\x1.3&1.8\x10$^{50}$&0.7	&4.5\x10$^{48}$ &19	&19\\
\enddata
\end{deluxetable}

For all portions of the Arched Filaments, the available ionizing
photons from the cluster are able to account for the ionization of the
H II regions. In most cases, a substantial excess in the number of
available ionizing photons exists compared with the number implied by the radio continuum. If the assumed
ionizing flux of the cluster is correct, the excess ionizing flux
indicates that the distance
between the Arches cluster and the Arched Filaments must be larger than the projected
distance. By assuming that the actual number of photons required is
equal to the estimated intercepted photon flux, it is possible to
solve for the line-of-sight component of distance for each
portion of the ionized gas. These derived line-of-sight components are
listed in Table 9, and range from 7 to 23 pc, about twice as large as the 3-10 pc range of
projected distances. Of course, if we have
overestimated the Lyman continuum output of the Arches cluster, then
the deduced line-of-sight distances would be smaller. Figure 22a shows
a schematic of the main portion of the Arched Filaments at their
projected distances from the Arches cluster, with greyscale
representing the magnitude of the line-of-sight component.  
The line-of-sight distances to the cluster for the E1 filament have the largest
values (23 pc for E1N and 19 pc for E1), whereas the estimated
distances to the southern portions of the
W1 and W2 filaments are only 7-9 pc along the line of sight to the cluster. 
With such large total distances between the cluster and ionized gas
(11-24 pc), the uniformity of the physical conditions observed across
the H II complex in these H92$\alpha$ line and in previous FIR studies
can be understood (Erickson et al. 1991; Colgan et al. 1996).

Using the derived line-of-sight distance between each filament and the
cluster, we can construct a schematic of a possible arrangement of
the filaments and the cluster. Figure 22b shows a view from above
(looking down the long axis of the cloud) of the $-$30 \kms~cloud and
the ionized filaments, which lie on its edge, and indicates the
position of the true observer. This schematic assumes that the ionized
filaments are on the near side of the molecular cloud (see $\S$5.1),
and that they are all located behind the Arches cluster. 
In order to produce the narrow, adjacent filaments in the Arched
Filaments, one may imagine a ``finger-like'' or columnar morphology
of the underlying molecular gas, such that the dense edges are
ionized by the Arches cluster. However, the relationship between the ionized and molecular gas in the
Arched Filaments is complex and Figure 22b is likely to be an
oversimplification of the system. A more detailed comparison between
the ionized and molecular gas in this region will be presented in Paper II
(Lang et al., in prep.). In addition, velocity information on the stars in the
Arches cluster will provide valuable information on the relative
geometry of the stars, ionized, and molecular gas and will be
available in the near future. 

\subsection{Comparison with Giant H II Regions NGC 3603 and 30 Dor}

To better understand the ionization of the Arched Filaments, it is
useful to compare its environment to the nebular structures which are
ionized by similarly powerful clusters: NGC 3603 and 30 Doradus in the
Large Magellanic Cloud. NGC 3603 is the largest optically
visible H II region in the Galaxy with a nebular diameter of \ab6 pc
and a ionized gas mass of \ab10$^4$ M$_{\sun}$ (Goss \&
Radhakrishnan 1969; Melnick, Tapia \& Terlevich 1989). The ionizing
cluster embedded in NGC 3603 contains at
least 20 O-stars and several Wolf-Rayet stellar types (Drissen et al. 1995). The R136 cluster which lies at the core
of the giant extragalactic H II region 30 Doradus contains more than a
dozen Wolf-Rayet stars, and \ab150 O-stars (Hunter et al. 1995). The
nebula in 30 Dor has a linear extent of more than 180 pc, but the
structures in the central 40 pc or so have been the subject of several recent studies (Hunter et al. 1995; Scowen et al. 1998). The Arched Filament nebula,
with a linear size of \ab20 pc and an ionizing cluster consisting of
$>$ 150 O-stars has similar characteristics to the clusters and nebulae of NGC 3603 and 30 Dor. In terms of the ionizing clusters, Figer et al. (1999) report that the total cluster mass of the Arches (10$^{4}$ M$_{\sun}$) is larger than that of NGC
3603, and comparable to that of the R136 cluster in the 30 Dor region. Similarly, the Arches, NGC 3603 and R136 clusters are relatively young, with estimated ages of 2, 4, and 2.5 Myr, respectively (Figer et al. 1999; Drissen et al. 1995; Hunter 1995).  The ionizing fluxes from these sources are also comparable: the
core of NGC 3603 (HD97950) produces \ab10$^{51}$ photons s$^{-1}$
(Drissen et al. 1995), the
R136 cluster ionizing flux is \ab2\x10$^{51}$ photons s$^{-1}$ (Hunter
et al. 1995), both
similar in value to the estimated 4\x10$^{51}$ photons s$^{-1}$ for the
Arches cluster (Morris et al. 2000). The nebular structure of NGC
3603 and 30 Dor have been the subject of many optical studies over the
years, and the kinematics of both regions are very complex (Clayton
1986, 1988, 1990; Hunter et al. 1995). 

However, there is an important difference between the Arched Filaments and these other complexes. In the case of the Arched Filament complex, the gravitational shearing forces on a molecular cloud located only 30 pc from the Galactic center are strong enough to either substantially displace the cluster from its natal cloud (with the help of stellar winds, supernovae, or radiation pressure) or to signficantly tidally disrupt the cloud over the lifetime of the cluster (2 Myr) so that it is no longer recognizable as a coherent cloud structure (see also $\S$5.1). Therefore, we cannot assume that the cluster formed out of this molecular cloud; consequently, the cluster is not likely to be ionizing away its natal material as is presumably occuring in NGC 3603 and 30 Dor. Instead, the $-$30 \kms~cloud is by chance passing by the cluster and as a consequence, is being strongly ionized. Although the relationship of the components of these stellar and nebular complexes may be different, the comparisons between the morphology and energetics of the Arched Filaments with 30 Dor and NGC 3603  can be quite instructive.

As in most H II regions, the ionized, molecular and stellar components in NGC 3603 are physically related. Clayton 
(1990) determined from optical nebular lines that the motions of the
gas in the core of NGC 3603 can be attributed to a number of
stellar-wind driven bubbles that have cleared out an area of the
surrounding interstellar medium around the cluster. Melnick et
al. (1989) also find that the spatial distribution of O-stars, ionized
gas and dust provide evidence that there is a cavity surrounding the
cluster. The radio recombination line study of NGC 3603 show kinematics consistent with at least one expanding
shell conicident with a faint optical wind bubble (DePree et
al. 1999).  In 30 Doradus, Chu \& Kennicutt (1994) identify a large
number of optical expanding shells, ranging in size from 1-100 pc.
Recent high-resolution HST optical images of the core
of the 30 Dor nebula in H\al~and other optical emission lines reveal a cavity around the stellar cluster, defined by a partial circular ring of emission at a radius of 5 pc from the cluster. The nebulosity outside of this cavity
consists of arcs and filamentary ridges up to 20 pc in extent, with extremely edge-brightened morphology (Scowen et al. 1998). 
The ionized gas in 30 Dor detected in the H90\al, H92\al, and H109\al~radio recombination lines shows a concentration
of emission along the western side, which defines a semi-circular
cavity around the stellar cluster. In addition, a double-peaked line
profile is observed, coincident in its velocity separation and
position with a known shell structure (Peck et al. 1997). 

As 30 Dor and NGC 3603 have similarly massive ionizing clusters at their cores, we might expect to detect shell-like structures in the vicinity of the Arches cluster. In addition, a cavity surrounding the Arches cluster might also be expected to be produced as the cluster begins to ionize and clear out lingering interstellar material. However, the velocity field of the H92\al~line observations does not
reveal any expanding shell structures, and double-peaked profiles are
observed in only two small regions. The Arched Filaments form a series of
concentric, partial rings of emission, centered roughly upon the Arches
Cluster, which could have an origin in a stellar wind-driven bubble structure. The narrow, curved arcs that make up the Arched Filaments extend over a range of scales from 2 to 15 pc, and at several positions (most notably in W1 in Figure 5), the ionized gas appears edge-brightened. The Eastern set of concentric rings is absent, indicating that the Arched Filaments H II complex is density-bounded on one side, and ionization-bounded on the other side, where the $-$30 \kms~molecular cloud is present (Serabyn \& \gusten~1987). 
Several bubbles in the Galactic center have already been detected in the mid-IR, the most prominent
of which is suggestively centered in the vicinity of the Radio Arc 
(Egan et al. 1998). It is likely to have been created by the Quintuplet cluster, an apparently older version of the Arches cluster (Kim et al. 1999). Correspondances between mid-IR features and those
at other wavelengths remain to be investigated. In addition, the large velocity gradients in a direction
perpendicular to the E1 and E2 filaments are not consistent with the
kinematics of the proposed molecular cloud orbit (see $\S$5.2). These anomolous
gas motions could be due to an interaction between the ionized gas and
the collective stellar winds from the Arches cluster. 

An important contribution to the overall energetics in the nebular gas in both
NGC 3603 and 30 Dor has been attributed to kinetic energy deposited to the gas from the stellar winds of the cluster
stars. In the case of NGC 3603, \ab3\x10$^{38}$ erg s$^{-1}$ is
released, and for 30 Dor, a wind luminosity of \ab1\x10$^{39}$ erg
s$^{-1}$ is estimated, which can produce wind-blown bubbles for NGC
3603 and 30 Dor with radii of 20 pc and 190 pc, respectively (Drissen et al. 1995; Hunter et al. 1995).
Over the lifetimes of these clusters (4 and 2.5 Myr respectively), the
mechanical energies released by the clusters are substantial (\ab10$^{52}$ ergs).
A portion of this energy goes into clearing out and heating the
interstellar gas to very high temperatures
(10$^7$-10$^8$ K) through stellar winds. In 30 Dor, ROSAT observations revealed that X-ray
emission arises from ``blister-shaped'' regions
closely correlated with loops of ionized gas (Wang 1999).
Canto, Raga \& Rodriguez (2000) predict that diffuse X-ray emission arising from
the combined stellar winds of the Arches cluster members (a ``cluster
wind'') should be detectable with sensitive X-ray observations. These authors predict that the Arches cluster wind luminosity 
should be \ab1\x10$^{38}$ erg s$^{-1}$, comparable to the wind
luminosities of the NGC 3603 and 30 Dor clusters. Such a wind is capable of imparting a significant amount of energy to the H II and molecular gas
in the Arched Filaments in the form of peculiar gas motions and shell-like
morphological structures. Thus, it is curious that shell structures are not more in evidence in the vicinity of the Arched Filaments. 

\section{Conclusions}

A VLA H92$\alpha$ recombination line and continuum study at 8.3 GHz of the kinematic and ionization properties of the Arched Filaments has been carried out. The following conclusions have been made:

(1) The 8.3 GHz continuum emission strongly resembles the 1.4 and 4.8
    GHz continuum images of the Arched Filaments from MYZ. The high resolution
    (2\farcs26 \x~1\farcs58, PA=64\fdg2) continuum image shows that
    the filaments have a remarkably uneven brightness with prominent
    ridges defining the edges of filaments.   

(2) The H92$\alpha$ recombination lines were imaged with a resolution
    of 12\farcs8 \x~8\farcs10, PA=1\arcdeg~and fit with
    single-component Gaussian models across the entire Arched Filaments. The H92\al~line properties (line-to-continuum ratios \ab0.1,
    $\Delta$V \ab28 \kms, average \T\ab6200 K)
    are found to be consistent with those for other Galactic center H II
    regions which have been photoionized by massive stars. 

(3) Narrow lines ($\Delta$V $<$ 20 \kms) are found in several regions
    of the Eastern Arched Filaments and these narrow lines place upper
    limits on the electron temperatures of 5400\p400 K, 6600\p400 K,
    and 6000\p600 K in three regions in the E1 and E2 filaments. 

(4) There are extremely large velocity gradients along each of the Arched Filaments. The
    sense of the velocity gradients in the W1, W2, and G0.10+0.02
    portions is increasingly negative velocities southward along the
    filaments, whereas in E1 and E2, the velocity gradients are not as
    organized, and in some cases have the opposite sense. The
    magnitudes of the velocity gradients range from 2$-$7 \kms~\pc~in the
    W1 and W2 filaments, and exceed 10 \kms~in the Eastern
    filaments. These magnitudes represent some of the largest
    gradients observed outside of the ionized gas streamers
    surrounding the black hole within a parsec of SgrA$^*$.

(5) High resolution H92\al~line observations (3\farcs61 \x~2\farcs66)
    were carried out for the SW region of the Arched Filaments. The
    recombination line properties are consistent with the lower
    resolution results. There do not appear to be any significant
    discontinuities in the velocity field, line widths, or
    line-to-continuum ratios in the region of the W1 filament where
    the Northern Thread NTF crosses in projection. Thus, the nature of
    a physical interaction, if any, remains unclear.

(6) The kinematics of the Arched Filaments are complex but can be broadly
    understood in terms of the orbit of the underlying molecular cloud
    around the Galactic center. The sense and magnitudes of the velocity gradient are consistent with the cloud residing on an x$_2$ orbit (a non-circular orbit family set up in response to the Galaxy's stellar bar) or to a radially infalling cloud, or to some combination of the two. Retrograde circular motion about the center cannot explain the observed kinematics and can be ruled out. 

(7) The Arched Filaments appear to be completely photoionized by the massive
    stars in the nearby Arches cluster. The cluster is
    likely to have a line-of-sight distance from the ionized filaments
    ranging from 7 to 23 pc, about a factor of two larger than the projected displacements. With a source of ionization located such large distances, the
    uniformity of the the physical conditions across the Arched
    Filaments can be understood. 

\section{Acknowledgements}
We thank Debra Shepherd for help with the mosaicking and Marc
Verheijen for help with GIPSY and the construction of the position-velocity
diagrams. We also acknowledge Angela Cotera for useful discussions regarding the ionization of the Arched Filaments, and Jun-Hui Zhao for making a
preliminary study of the H92$\alpha$ lines in the Arched Filaments.

\end{document}